\documentclass{emulateapj}   
\submitted{ApJ, accepted}

\usepackage{rotating}

\def\ltsima{$\; \buildrel < \over \sim \;$}
\def\simlt{\lower.5ex\hbox{\ltsima}} 
\def\gtsima{$\; \buildrel > \over \sim \;$}
\def\simgt{\lower.5ex\hbox{\gtsima}} 
\def\arcsec{\hbox{$^{\prime\prime}$}}
\def\deg{\hbox{$^\circ$}}
\def\cgsflux{erg s$^{-1}$ cm$^{-2}$}

\def\r95{$r_{\rm 95}$}

\def\Fermi{\textit{Fermi}}
\def\Suzaku{\textit{Suzaku}}
\def\Swift{\textit{Swift}}
\def\Chandra{\textit{Chandra}}
\def\XMM{\textit{XMM}}

\def\aJ1311{0FGL~J1311.9$-$3419}
\def\bJ1653{0FGL~J1653.4$-$0200}
\def\cJ2214{0FGL~J2214.8+3002}
\def\dJ2241{0FGL~J2241.7$-$5239}
\def\eJ2339{0FGL~J2339.8$-$0530}

\def\wJ1311a{CXOU~J131145.7$-$343030}
\def\xJ1311b{CXOU~J131147.0$-$343205}
\def\yJ1653a{CXOU~J165338.0$-$015836}
\def\zJ1653b{CXOU~J165341.4$-$015927}

\shorttitle{\Chandra\ Observations of Bright \Fermi\ Unidentified Sources}
\shortauthors{Cheung et al.}

\begin{document}

\title{\Chandra\ X-ray Observations of the Two Brightest Unidentified
High Galactic Latitude \Fermi-LAT $\gamma$-ray Sources}

\author{
C.~C.~Cheung\altaffilmark{1,2}, 
D.~Donato\altaffilmark{3,4},
N.~Gehrels\altaffilmark{2},
K.~V.~Sokolovsky\altaffilmark{5,6}, 
M.~Giroletti\altaffilmark{7}
}

\altaffiltext{1}{National Research Council Research Associate, National 
Academy of Sciences, Washington, DC 20001, resident at Naval Research 
Laboratory, Washington, DC 20375, USA. Teddy.Cheung.ctr@nrl.navy.mil}
\altaffiltext{2}{NASA Goddard Space Flight Center, Greenbelt, MD 20771,
USA.}
\altaffiltext{3}{Center for Research and Exploration in Space Science 
and Technology (CRESST) and NASA Goddard Space Flight Center, Greenbelt, 
MD 20771, USA. Davide.Donato-1@nasa.gov}
\altaffiltext{4}{Department of Physics and Department of Astronomy, 
University of Maryland, College Park, MD 20742, USA.}
\altaffiltext{5}{Astro Space Center of the Lebedev Physical Institute, 
117810 Moscow, Russia.}
\altaffiltext{6}{Sternberg Astronomical Institute of the Moscow State 
University, 119992 Moscow, Russia.}
\altaffiltext{7}{INAF Istituto di Radioastronomia, 40129 Bologna, Italy.}

\begin{abstract}

We present \Chandra\ ACIS-I X-ray observations of \aJ1311 and \bJ1653, 
the two brightest high Galactic latitude ($|b|>$10\deg) $\gamma$-ray 
sources from the 3 month \Fermi-LAT bright source list that are still 
unidentified. Both were also detected previously by EGRET, and despite 
dedicated multi-wavelength follow-up, they are still not associated with 
established classes of $\gamma$-ray emitters like pulsars or radio-loud 
active galactic nuclei. X-ray sources found in the ACIS-I fields of view 
are catalogued, and their basic properties are determined. These are 
discussed as candidate counterparts to \aJ1311\ and \bJ1653, with 
particular emphasis on the brightest of the 9 and 13 \Chandra\ sources 
detected within the respective \Fermi-LAT 95$\%$ confidence regions. 
Further follow-up studies, including optical photometric and 
spectroscopic observations, are necessary to identify these X-ray 
candidate counterparts in order to ultimately reveal the nature of these 
enigmatic $\gamma$-ray objects.

\end{abstract}


\section{Introduction\label{sec-intro}}

Since its launch in 2008, the \Fermi\ Large Area Telescope 
\citep[LAT;][]{atw09} has enabled substantial progress in our 
understanding of the high-energy (HE; $>$100 MeV) $\gamma$-ray Universe. 
A long-standing problem in the field has been in the secure 
identification of discrete HE $\gamma$-ray sources, with most objects 
detected previously by COS B \citep{swa81} and EGRET 
\citep{har99,cas08}, remaining unidentified prior to the present 
\Fermi-era. These sources have eluded identification mainly due to high 
source confusion in the poorly localized $\gamma$-ray regions with 
typical 95$\%$ confidence radii, \r95 $\sim 0.4\deg-0.7\deg$ in the 3rd 
EGRET catalog \citep[3EG;][]{har99}. Also, for variable HE emitters 
\citep[e.g.,][]{tav97,nol03}, there was a lack of prompt response to the 
$\gamma$-ray events, with much of the multi-wavelength follow-up work 
pursued many years later \citep[e.g.,][]{par08}.

Multi-wavelength follow-up observations of unidentified 3EG sources 
attracted much effort, but met with mixed success \citep[see][for a 
summary]{muk04}. The LAT's all-sky monitoring capability (20$\%$ of the 
sky at all times), its increased sensitivity ($>$10$\times$ better than 
EGRET), and dramatically improved localizations over EGRET, has enabled 
many of the unidentified EGRET sources to be successfully associated 
with lower-energy counterparts. The dominant $\gamma$-ray emitting 
population consists of radio-loud active galactic nuclei (AGN), 
including blazars and radio galaxies. Perhaps unexpectedly, a 
substantial population of new $\gamma$-ray pulsars have been identified 
($\gamma$-ray pulsations detected via a blind search or using known 
radio ephemerides), along with a handful of pulsar wind nebula, 
supernova remnants, and $\gamma$-ray binaries. While the population of 
EGRET unidentified sources has quickly diminished, the number of fainter 
unidentified \Fermi-LAT $\gamma$-ray objects has been increasing 
\citep[see][for a summary]{2fgl}.

\begin{table*}
\footnotesize
\begin{center}
\caption{}
\begin{tabular}{lccccccc}
\hline\hline
\multicolumn{1}{c}{Name} &
\multicolumn{2}{c}{Pointing Center} &
\multicolumn{2}{c}{LAT Centroid} &
\multicolumn{1}{c}{ObsID} &
\multicolumn{1}{c}{Start Time} &
\multicolumn{1}{c}{Net Exp.} \\
\cline{2-3}\cline{4-5}
\multicolumn{1}{c}{}&
\multicolumn{1}{c}{R.A.} &
\multicolumn{1}{c}{Decl.} &
\multicolumn{1}{c}{$l$ (deg)} &
\multicolumn{1}{c}{$b$ (deg)} &
\multicolumn{1}{c}{} &
\multicolumn{1}{c}{in 2010 (UT)} &
\multicolumn{1}{c}{(ks)}\\
\hline
 0FGL~J1311.9$-$3419 & 13 11 49.68 &--34 29 31.2 & 307.686 & +28.195 & 11790 & Mar 21 15:38:54 & 19.87 \\
 0FGL~J1653.4$-$0200 & 16 53 43.68 &--01 58 30.0 & 16.593 & +24.931 & 11787 & Jan 24 06:21:27 & 20.77 \\
\hline
\hline
\end{tabular}
\end{center}
\Chandra\ observational summary for the two targets. The
positions are the pointing centers (J2000.0 equinox) set at the time of 
the observations and the LAT centroids are the 2FGL catalog values in
Galactic coordinates. The \Chandra\ observation ID (ObsID), start time,
and net exposure (Net Exp.) are also provided.
\label{table-1}
\end{table*}

As part of a systematic study of \Fermi-LAT unidentified sources in 
X-rays, including \Suzaku\ \citep[e.g.,][]{mae11,tak12a}, \Swift\ 
\citep{fal11}, and \XMM-Newton \citep[e.g.,][]{wol10}, we obtained new 
\Chandra\ observations in cycle-11 covering the fields of five 
unidentified high Galactic latitude sources from the initial 3 month 
\Fermi-LAT bright source list \citep[0FGL;][]{bsl}. With its large field 
of view ($17' \times 17'$, sufficient to cover the LAT 95$\%$ confidence 
regions completely) and excellent sensitivity (5$\sigma$ flux 
sensitivity of $\sim 1.5 \times 10^{-14}$ \cgsflux, $0.5-8$ keV), 
\Chandra\ observations allow for a sensitive study of all X-ray sources 
within the LAT localization regions. For the three targets subsequently 
identified as pulsars, PSR~J2214+3002/\cJ2214 \citep{ran11} and 
PSR~J2241$-$5236/\dJ2241 \citep{kei11}, and a possible radio quiet 
millisecond pulsar (MSP) in \eJ2339\ \citep{kon12}, we reported the 
results of our \Chandra\ observations in those papers. Here, we present 
the results of the \Chandra\ study of the remaining two objects 
(\aJ1311\ and \bJ1653)\footnote{Throughout, we retain the 0FGL names 
although newer information, particularly from the 2FGL release, are 
used. For reference, the various \Fermi-LAT catalog names for the 
sources studied in this paper are: 0FGL~J1311.9$-$3419 = 
1FGL~J1311.7$-$3429 = 2FGL~J1311.7$-$3429, and 0FGL~J1653.4$-$0200 = 
1FGL~J1653.6$-$0158 = 2FGL~ J1653.6$-$0159.}. These \Fermi-LAT sources 
were detected previously by EGRET, as 3EG~J1314$-$3431/EGR~J1314$-$3417 
and 3EG~J1652$-$0223/EGR~J1653$-$0249 \citep{har99,cas08}, and are two 
of the brightest remaining unidentified sources from that era.

In the following, we describe the analysis of the \Chandra\ observations 
in Sec.~2, including the X-ray detection and 
characterization methods (Sec.~\ref{sec-fields}) and a more detailed 
analysis and discussion of individual X-ray sources found within and 
near the LAT localization regions (Sec.~\ref{sec-bright}).  Results from 
a search for positional matches with archival optical, near-infrared, 
mid-infrared, and radio catalogs are also summarized in 
Sec.~\ref{sec-bright}. We then discuss the general population of X-ray 
sources as potential counterparts to the $\gamma$-ray objects with 
particular emphasis on the aforementioned brightest \Chandra\ sources 
(Sec.~3), and conclude with a summary of the results 
(Sec.~4).

\section{\Chandra\ X-ray Observations\label{sec-chandra}}

We obtained \Chandra\ X-ray Observatory (CXO) observations of the two 
targets in early 2010 with $\sim$20 ks exposures each 
(Table~\ref{table-1}). At the time of the observational planning, the 
available LAT localization errors based on 3 months of data \citep[\r95 
= $12.2'$ and $9.5'$;][]{bsl} and from an internal \Fermi-LAT team 
analysis of 6 months of data (\r95 $\sim~4' - 5'$) were still relatively 
large. We thus opted for the larger field of view (FOV) provided by the 
ACIS-I detector ($\sim 17' \times 17'$) over ACIS-S ($\sim 8' \times 
8'$) to ensure sufficient coverage of the LAT error regions. With the 
currently available analysis of 2-years of LAT data from the 2FGL 
catalog \citep{2fgl}, the localization errors have since improved to 
\r95 $\simeq~2.0'$ for \aJ1311\ (Fig.~\ref{image1}) and $\simeq~3.6'$ 
for \bJ1653\ (Fig.~\ref{image2}), and we utilize these values for the 
remainder of this paper.

The observations were obtained in FAINT mode at the nominal ACIS-I 
aim-point centered toward the corner of one (I3) of its four $2 \times 
2$ arrayed CCDs. The default dithering mode of \Chandra\ observations 
enables exposure in the gaps between the CCDs, but obviously with 
decreased effective area. In this context, the brightest \Chandra\ 
source (\wJ1311a) within the 2FGL error ellipse of \aJ1311\ fell in such 
a gap. This resulted in a smaller effective area at its position 
($\sim$215 cm$^{2}$ at 1.5 keV compared to the maximum value of 
$\sim$500 cm$^{2}$ in the field; see Sec.~\ref{sec-fields} and 
Appendix), thus decreasing its observed count rate and statistics for 
temporal and spectral analysis (Sec.~\ref{sec-bright}).

\subsection{Analysis of the ACIS-I Fields\label{sec-fields}}

For the analysis, we used the CIAO software \citep{fru06} version 4.2, 
with updated calibration files from CALDB version 4.2.2, and the {\tt 
ACIS Extract} \citep{bro10} software version 2010-02-26. The data were 
downloaded from the \Chandra\ data archive and reprocessed from the 
level 1 (evt1) files following the standard CIAO procedure: (1) we 
created a new ACIS bad pixel file using {\tt acis$\_$run$\_$hotpix}, (2) 
we ran {\tt acis$\_$process$\_$events} to create a new evt1 file with 
the calibration applied, (3) we identified afterglows (cosmic-ray 
residual events) in the CCDs using {\tt acis$\_$detect$\_$afterglow}, 
and (4) we filtered the data on grade, status, and good time, generating 
two level 2 (evt2) files following the recipe of \citet[][Appendix A 
therein]{bro10}. In the last step, a more aggressive afterglow cleaning 
was applied to one evt2 file for the source detection, while less 
aggressive afterglow cleaning was applied to the second for the actual 
source extraction and analysis.  We checked the evt2 files for 
background flares and none were detected.

To search for candidate X-ray sources over the four ACIS-I chips 
covering each field, we ran {\tt wavdetect} in the full energy band 
($0.5-8$ keV) and in the two sub-bands, $0.5-2$ keV (soft) and $2-8$ keV 
(hard), using images binned by 0.5, 1, and 2 ACIS pixels (0.492\arcsec\ 
pixel$^{-1}$). The resultant source lists found in each run were merged 
to produce a single master list. We then used {\tt ACIS Extract} to 
excise insignificant candidate sources, to scan for sources containing 
afterglow events, to refine the source positions, and to perform 
photometric, spectral, and temporal analysis.

\begin{figure}
  \begin{center}
    \includegraphics[width=8.5cm]{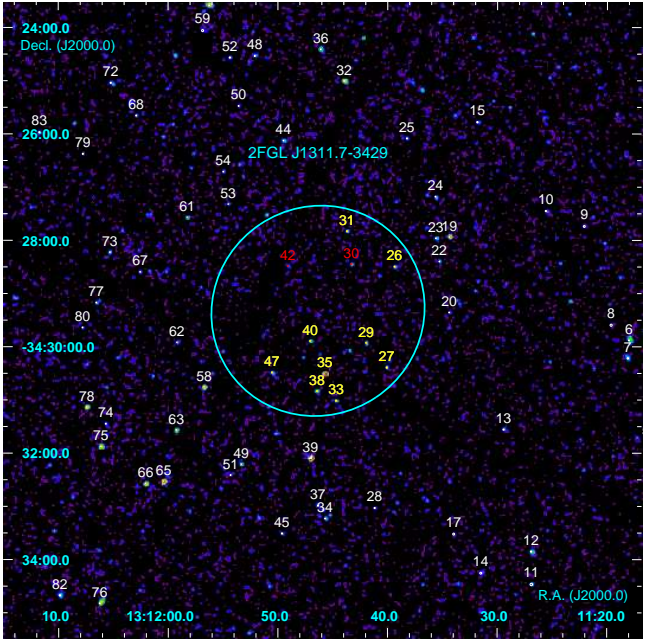}
  \end{center}
\caption{\Chandra\ ACIS-I image of the central $12' \times 12'$ field of
\aJ1311\ with 2FGL 95$\%$ confidence error ellipse plotted (cyan). The
image was binned by $2 \times 2$ pixels (0.492\arcsec\ pixel$^{-1}$) and
Gaussian smoothed with a kernel radius of 3 pixels. X-ray sources lying
within this FOV are marked with their corresponding catalog numbers and
are divided into sources outside (white) and inside (yellow =
detections, red = tentative sources) the 2FGL ellipse.
}
\label{image1}
\end{figure}

\begin{figure}
  \begin{center}
    \includegraphics[width=8.5cm]{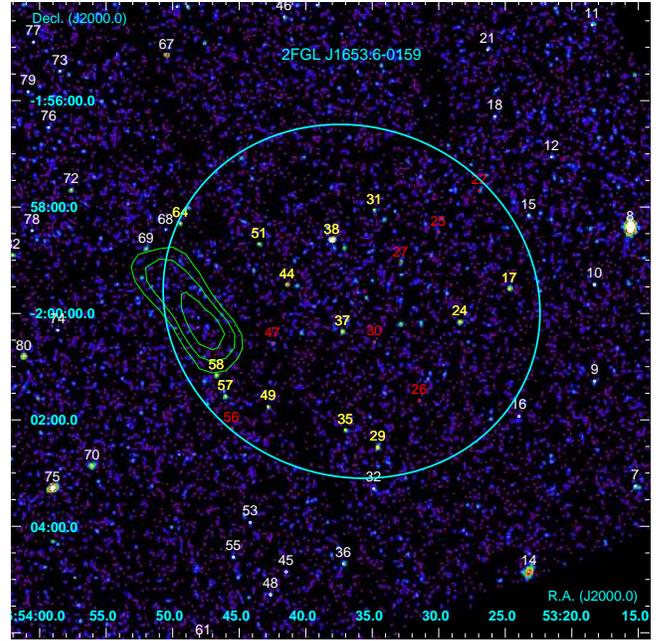}
  \end{center}
\caption{As in Fig.~\ref{image1}, but for \bJ1653. In addition, the NVSS
1.4 GHz radio contours for NVSS~J165348.44$-$015958.7 (discussed in   
Sec.~3) are indicated in green contours (levels of
1.5, 2.1, 3.0 mJy beam$^{-1}$; 45\arcsec\ beam).
}
\label{image2}
\end{figure}

For both targets, 97 total sources were found in our analysis of each of 
the two ACIS-I fields and the complete source lists are tabulated in 
Appendix A. Included in the tables are the source identifiers in the 
form of a catalog number (N; ordered from increasing R.A.) and a 
CXOU~J2000.0 coordinate based name, source positions (R.A.~and Decl.~in 
J2000.0) and corresponding errors ($r$, statistical 
only)\footnote{Comparing our list of X-ray sources with USNO~B1.0 
catalog matches, we found a negligible offset in the \aJ1311\ field 
positions, but a systematic offset of $0.52\arcsec \pm 0.39\arcsec$ in 
the \bJ1653\ field positions (see Appendix~B for details). This is 
within the expected systematic uncertainty of 0.6\arcsec\ in \Chandra's 
astrometry (see: http://cxc.harvard.edu/cal/ASPECT/celmon/), which when 
considering absolute positional errors, should be added in quadrature 
with the statistical values.}. The catalog numbers of the X-ray sources 
in the central $12' \times 12'$ portion of the \Chandra\ images are 
marked in Fig.~\ref{image1} (\aJ1311) and Fig.~\ref{image2} (\bJ1653). 
We also provide the source distance from the \Chandra\ aim-point, the 
effective area at 1.5 keV, net X-ray counts in each of the $0.5-8$ keV, 
$0.5-2$ keV, and $2-8$ keV energy ranges, along with their respective 
source significances in standard deviations ($\sigma$) and probabilities 
($P_{\rm B}$) determined by {\tt ACIS Extract} for the null source 
hypothesis. To gauge variability, a one-sample Kolmogorov-Smirnov (KS) 
statistic was used to calculate the probability, $P_{\rm KS}$, for the 
null hypothesis of a uniform flux for all sources. Following 
\citet{bro10}, we identify five X-ray sources in each of the two fields 
to be possibly variable ($0.005 < P_{\rm KS} < 0.05$), and the remaining 
show no evidence for variability ($P_{\rm KS} > 0.05$) -- see Appendix~A 
for details. One of the possibly variable sources is located within the 
2FGL 95$\%$ localization of \bJ1653\ (Sec.~\ref{sec-bright}), while the 
rest are positioned outside of the 95$\%$ confidence ellipses.

\begin{table*}
\footnotesize
\begin{center}
\caption{}
    \tabcolsep 4.8pt
\begin{tabular}{lccccccc}
\hline\hline
\multicolumn{1}{c}{N} &
\multicolumn{1}{c}{CXOU Name} &
\multicolumn{1}{c}{$r$} &
\multicolumn{1}{c}{Counts} &
\multicolumn{1}{c}{$\sigma$} &
\multicolumn{1}{c}{Counts} &
\multicolumn{1}{c}{Counts} &
\multicolumn{1}{c}{$F_{\rm 0.5-8~keV}$} \\
\cline{4-5}
\multicolumn{1}{c}{} &
\multicolumn{1}{c}{} &
\multicolumn{1}{c}{($\arcsec$)} &
\multicolumn{2}{c}{($0.5-8$ keV)} &     
\multicolumn{1}{c}{($0.5-2$ keV)} &
\multicolumn{1}{c}{($2-8$ keV)} &
\multicolumn{1}{c}{}\\
\hline
\multicolumn{7}{c}{\aJ1311} \\
\hline
 26 & J131139.3-342829 & 0.23 &    5.9 (+3.6/-2.4) & 1.6 &    4.0 (+3.2/-1.9) &    1.9 (+2.7/-1.3) &  0.3 \\ 
 27 & J131140.0-343023 & 0.30 &    3.9 (+3.2/-1.9) & 1.2 &    2.0 (+2.7/-1.3) &    1.9 (+2.7/-1.3) &  0.2 \\ 
 29 & J131141.9-342955 & 0.19 &    6.9 (+3.8/-2.6) & 1.8 &    1.0 (+2.3/-0.8) &    6.0 (+3.6/-2.4) &  0.4 \\ 
 31 & J131143.7-342749 & 0.26 &    3.9 (+3.2/-1.9) & 1.2 &    3.0 (+2.9/-1.6) &    0.9 (+2.3/-0.8) &  0.2 \\ 
 33 & J131144.6-343100 & 0.24 &    4.9 (+3.4/-2.2) & 1.4 &    2.0 (+2.7/-1.3) &    3.0 (+2.9/-1.6) &  0.3 \\ 
 35 & J131145.7-343030 & 0.07 &   54.0 (+8.4/-7.3) & 6.4 &   32.0 (+6.7/-5.6) &   22.0 (+5.8/-4.7) &  10.3 \\ 
 38 & J131146.4-343050 & 0.18 &    7.9 (+4.0/-2.8) & 2.0 &    2.0 (+2.7/-1.3) &    6.0 (+3.6/-2.4) &  0.4 \\ 
 40 & J131147.0-342953 & 0.21 &    4.0 (+3.2/-1.9) & 1.2 &    3.0 (+2.9/-1.6) &    1.0 (+2.3/-0.8) &  0.2 \\ 
 47 & J131150.5-343029 & 0.23 &    3.9 (+3.2/-1.9) & 1.2 &    3.0 (+2.9/-1.6) &    1.0 (+2.3/-0.8) &  0.2 \\ 
\hline
 30*& J131143.3-342826 & 0.27 &    2.9 (+2.9/-1.6) & 1.0 &   -0.0 (+1.9/-0.0) &    3.0 (+2.9/-1.6) &  0.2 \\ 
 42*& J131149.0-342829 & 0.23 &    3.0 (+2.9/-1.6) & 1.0 &    1.0 (+2.3/-0.8) &    2.0 (+2.7/-1.3) &  0.2 \\ 
\hline
\hline
\multicolumn{7}{c}{\bJ1653} \\
\hline
 17 & J165324.6-015932 & 0.40 &   10.5 (+4.4/-3.3) & 2.4 &    6.8 (+3.8/-2.6) &    3.7 (+3.2/-1.9) &  0.7 \\ 
 24 & J165328.4-020009 & 0.28 &   13.7 (+4.8/-3.7) & 2.8 &   11.9 (+4.6/-3.4) &    1.8 (+2.7/-1.3) &  0.8 \\ 
 29 & J165334.5-020230 & 0.38 &   13.5 (+4.8/-3.7) & 2.8 &    7.8 (+4.0/-2.8) &    5.7 (+3.6/-2.4) &  0.7 \\ 
 31 & J165334.8-015803 & 0.23 &    4.9 (+3.4/-2.2) & 1.4 &    3.0 (+2.9/-1.6) &    2.0 (+2.7/-1.3) &  0.3 \\ 
 35 & J165337.0-020211 & 0.49 &    5.6 (+3.6/-2.4) & 1.6 &    4.8 (+3.4/-2.2) &    0.8 (+2.3/-0.8) &  0.3 \\ 
 37 & J165337.2-020020 & 0.18 &   11.9 (+4.6/-3.4) & 2.6 &    5.0 (+3.4/-2.2) &    6.9 (+3.8/-2.6) &  0.6 \\ 
 38 & J165338.0-015836 & 0.03 &  306.9 (+18.5/-17.5) & 16.6 &  198.0 (+15.1/-14.1) &  109.0 (+11.5/-10.4) & 23.1 \\ 
 44 & J165341.4-015927 & 0.10 &   18.9 (+5.4/-4.3) & 3.5 &   17.0 (+5.2/-4.1) &    2.0 (+2.7/-1.3) &  0.7 \\ 
 49 & J165342.8-020144 & 0.38 &    5.8 (+3.6/-2.4) & 1.6 &    5.9 (+3.6/-2.4) &   -0.1 (+1.9/-0.0) &  0.3 \\ 
 51 & J165343.4-015841 & 0.16 &    5.9 (+3.6/-2.4) & 1.6 &    6.0 (+3.6/-2.4) &   -0.0 (+1.9/-0.0) &  0.3 \\ 
 57 & J165346.0-020133 & 0.28 &    8.8 (+4.1/-2.9) & 2.1 &    6.9 (+3.8/-2.6) &    1.9 (+2.7/-1.3) &  0.5 \\ 
 58 & J165346.7-020109 & 0.21 &   11.8 (+4.6/-3.4) & 2.6 &    5.9 (+3.6/-2.4) &    5.9 (+3.6/-2.4) &  0.6 \\ 
 64 & J165349.4-015818 & 0.20 &    6.0 (+3.6/-2.4) & 1.7 &    1.0 (+2.3/-0.8) &    5.0 (+3.4/-2.2) &  0.5 \\ 
\hline
 22*& J165326.9-015741 & 0.61 &    2.7 (+2.9/-1.6) & 0.9 &    1.9 (+2.7/-1.3) &    0.8 (+2.3/-0.8) &  0.2 \\ 
 25*& J165330.0-015827 & 0.45 &    2.8 (+2.9/-1.6) & 1.0 &    1.9 (+2.7/-1.3) &    0.9 (+2.3/-0.8) &  0.2 \\ 
 26*& J165331.4-020137 & 0.73 &    2.6 (+2.9/-1.6) & 0.9 &   -0.2 (+1.9/-0.0) &    2.8 (+2.9/-1.6) &  0.1 \\ 
 27*& J165332.8-015902 & 0.37 &    2.9 (+2.9/-1.6) & 1.0 &    0.9 (+2.3/-0.8) &    1.9 (+2.7/-1.3) &  0.2 \\ 
 30*& J165334.8-020031 & 0.43 &    2.9 (+2.9/-1.6) & 1.0 &   -0.1 (+1.9/-0.0) &    2.9 (+2.9/-1.6) &  0.2 \\ 
 47*& J165342.4-020034 & 0.36 &    3.0 (+2.9/-1.6) & 1.0 &    2.0 (+2.7/-1.3) &    1.0 (+2.3/-0.8) &  0.3 \\ 
 56*& J165345.5-020209 & 0.63 &    2.7 (+2.9/-1.6) & 0.9 &    0.8 (+2.3/-0.8) &    1.8 (+2.7/-1.3) &  0.1 \\ 
\hline
\hline
\end{tabular}
\end{center}
Detected and tentatively detected (catalog numbers, N, marked with asterisk) 
\Chandra\ X-ray sources within the 2FGL error ellipses of 
\aJ1311\ and \bJ1653. Listed are the J2000.0 coordinate based names, 
error radius ($r$), net counts in each of the three defined energy 
bands, and significance ($\sigma$) in the $0.5-8$ keV band. The $0.5-8$ 
keV fluxes ($F_{\rm 0.5-8~keV}$; in units of $10^{-14}$ \cgsflux) are 
from the spectral fits for the brightest sources 
(Tables~\ref{table-j1311}~\&~\ref{table-j1653}), while the estimates for 
the remaining fainter sources are described in Sec.~\ref{sec-fields}.
\label{table-2}
\end{table*}

In order to discuss the most likely candidate X-ray counterparts to the 
unidentified $\gamma$-ray sources, we consider only the \Chandra\ 
detected sources within the 2FGL 95$\%$ confidence error ellipses. As 
the LAT localization errors for the two targets are different, we set 
separate detection thresholds, $P_{\rm B}$ $<5.4 \times 10^{-6}$ for 
\aJ1311\ and $<1.8 \times 10^{-6}$ for \bJ1653, in any of the three 
defined X-ray energy bands. This corresponds to less than one false 
positive source due to background fluctuations within each 2FGL ellipse. 
For \aJ1311\ and \bJ1653, we detected 9 and 13 such X-ray sources, while 
an additional two and seven tentative ones (those that did not pass the 
detection threshold) were found, respectively. In Table~\ref{table-2}, 
we provide selected information for these X-ray sources taken from 
Appendix~A. It is apparent from the table that our choices for the 
probability thresholds divided detected X-ray sources into ones with 
$\simgt 4$ net counts and $>1.0 \sigma$ in the full ($0.5-8$ keV) band, 
while those with $\sim 3$ net counts and $\simlt 1.0 \sigma$ were deemed 
tentative. The detected sources are indicated by yellow markers in 
Figures~\ref{image1}~\&~\ref{image2}, while the tentative ones by red 
markers (the latter are no longer discussed in the main text of the 
paper). In Table~\ref{table-2}, we also provide $0.5-8$ keV fluxes. For 
the brightest sources (\wJ1311a\ = N35, \yJ1653a\ = N38, and \zJ1653b\ = 
N44), the fluxes are from detailed spectral fits 
(Sec.~\ref{sec-bright}), while for the remaining fainter sources, we 
estimated fluxes using exposure corrected count rates and a conversion 
$10^{-3}$ counts s$^{-1}$ = $10^{-14}$ \cgsflux, calculated from 
PIMMS\footnote{http://cxc.harvard.edu/toolkit/pimms.jsp} assuming a 
single power-law with photon index, $\Gamma = 2$.

\subsection{Further X-ray Analysis for Selected Sources and 
Multi-wavelength Counterpart Matches\label{sec-bright}}

\begin{figure}
  \begin{center}
    \includegraphics[width=9.5cm]{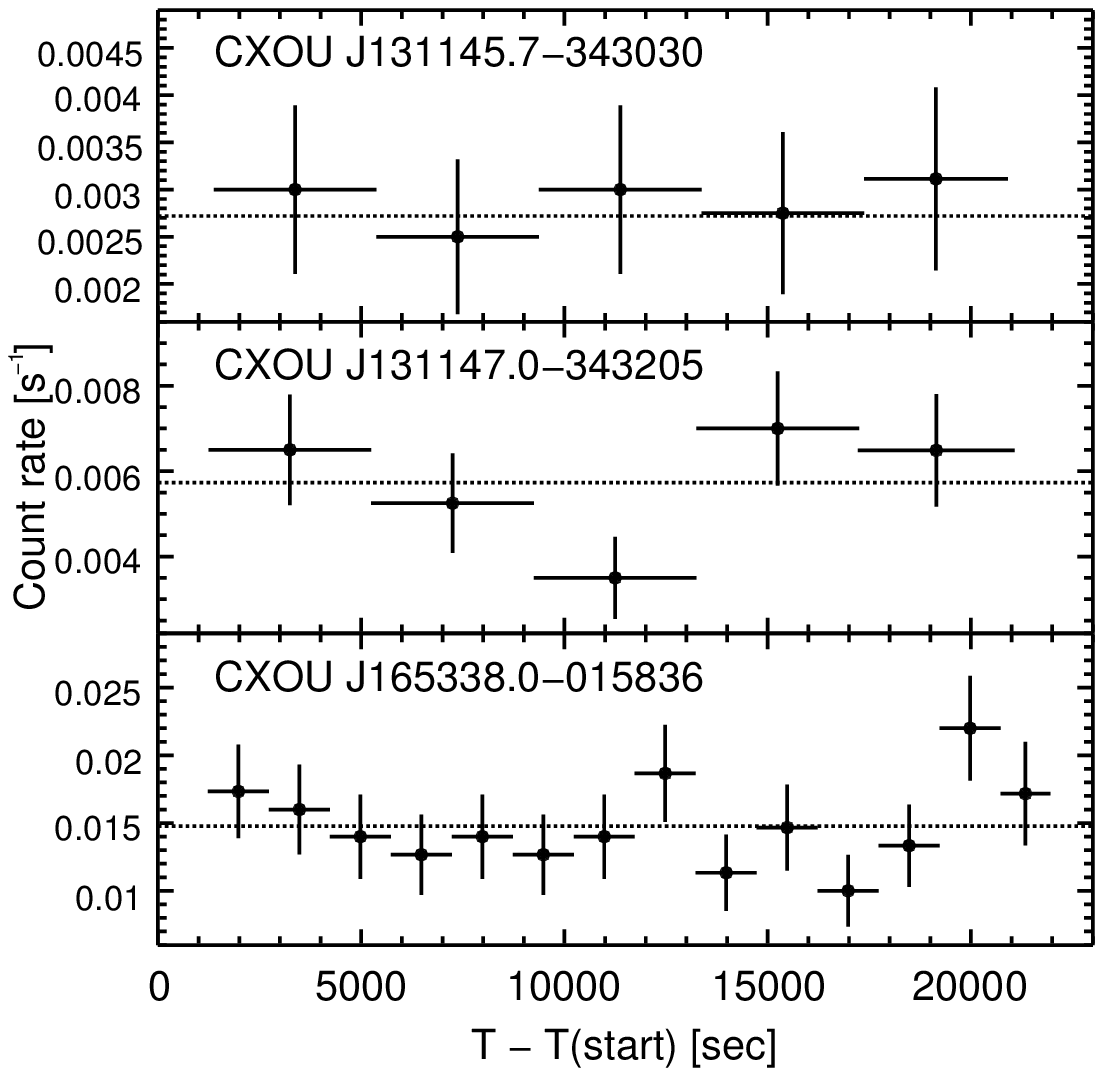}
  \end{center}
\vspace{-0.1in}
\caption{\Chandra\ ACIS-I ($0.5-8$ keV) lightcurves of the brightest
X-ray source within (\wJ1311a, top) and just outside and to the south
(\xJ1311b, middle) of the 2FGL error ellipse of \aJ1311, and the
brightest X-ray source within the 2FGL error ellipses of \bJ1653\
(\yJ1653a, bottom). The data points are shown in time bins that are 4 ks
($\sim 10$ counts per bin), 4 ks ($\sim 20$ counts per bin), and 1.5 ks
($\sim 20$ counts per bin) long, respectively. The times are indicated
relative to the observation start times (Table~\ref{table-1}) and the
average count rates are indicated with horizontal dotted lines.
}
\label{lightcurve}
\end{figure}

Several X-ray sources have sufficient statistics to perform more 
detailed spectral and temporal analysis and these results are presented 
in the following (Sec.~\ref{sec-j1311}~\&~\ref{sec-j1653}). 
Specifically, we studied the brightest X-ray sources within the 2FGL 
error ellipses, namely, \wJ1311a\ (Fig.~\ref{image1}) and \yJ1653a\ 
(Fig.~\ref{image2}), with respective $0.5-8$ keV count rates of $(2.72 
\pm 0.37) \times 10^{-3}$ counts s$^{-1}$ and $(14.78 \pm 0.84) \times 
10^{-3}$ counts s$^{-1}$. As mentioned (Sec.~2), the 
former count rate is relatively low because the X-ray source fell on a 
gap between the ACIS-I CCDs. Another X-ray bright source (\wJ1311a; 
$(5.73 \pm 0.54) \times 10^{-3}$ counts s$^{-1}$) just outside and to 
the south of the 2FGL error ellipse (Fig.~\ref{image1}) was previously 
detected in a \Suzaku\ observation \citep{mae11} so is also discussed 
here. Lightcurves for these sources were generated 
(Fig.~\ref{lightcurve}), confirming the finding based on the KS-test 
(Sec.~\ref{sec-fields}) that there is no significant X-ray variability 
within the $\sim$20 ks observation span. Their X-ray spectra, along with 
that of the second brightest X-ray source within the 2FGL error ellipse 
of \bJ1653\ (\zJ1653b) were analyzed using {\tt XSPEC v12}, considering 
Galactic absorption ($N_{\rm H, Gal}$) values from \citet{kal05}. These 
X-ray spectral fitting results are summarized in 
Tables~\ref{table-j1311}~\&~\ref{table-j1653}.

From the KS-test applied to all the detected X-ray sources 
(Sec.~\ref{sec-fields}), only one source with possible variability is 
detected within a \Fermi-LAT localization ellipse and notes are provided 
on this object (CXOU~J165337.2$-$020020 in \bJ1653). Spectral results 
are additionally provided for the bright, prominent X-ray source, 
CXOU~J165315.6$-$015822 (seen toward the right edge of 
Fig.~\ref{image2}), found outside of its corresponding 2FGL localization 
ellipse. For the remaining sources, the statistics allow us to only 
broadly characterize the hardness or softness of their X-ray spectra.

The typical sub-arcsecond localizations provided by our \Chandra\ 
observations (Table~\ref{table-2} and Appendix~A) allow us to search 
confidently for optical, near-infrared, mid-infrared, and radio 
counterparts to the X-ray sources using the USNO~B1.0 \citep{mon03}, 
2MASS \citep{skr06}, WISE \citep{wri10} from the preliminary data 
release \citep[WISEP;][]{cut12}, and NVSS \citep{con98} catalogs. The 
results of the USNO~B1.0, 2MASS, and WISEP catalog matches for the 97 
X-ray sources detected in each of the two ACIS-I fields are presented in 
Appendix~B. Of the X-ray sources within the 2FGL 95$\%$ confidence error 
regions of \aJ1311\ and \bJ1653, and considering optical/infrared 
sources within 2\arcsec\ of the X-ray positions, we found 2/9 and 3/13 
of the detected sources with USNO~B1.0 counterparts, respectively. Of 
the \Chandra/USNO~B1.0 matches, there was one case in \aJ1311\ where a 
WISEP match was also found (but without an 2MASS counterpart), and two 
cases in \bJ1653\ with 2MASS and WISEP counterparts also. There were 
single detected X-ray sources within the two 2FGL error ellipses where 
we found possible matches with a WISEP source, but without counterparts 
in the USNO~B1.0 and 2MASS. Specific notes are provided for the 
optical/infrared matched \Chandra\ sources in the following subsections. 
None of the X-ray sources in the two ACIS-I fields had radio 
counterparts in the NVSS catalog (45\arcsec\ resolution at 1.4 GHz with 
typical sensitivity level of $\sim 2.5$ mJy) out to a 45\arcsec\ search 
radius. In fact, within the 2FGL 95$\%$ confidence ellipses of the two 
$\gamma$-ray objects, only one NVSS source is found (within \bJ1653; see 
Figure~\ref{image2}) and it is briefly discussed in 
Sec.~3.

\begin{table*}
\begin{center}
\caption{}
\begin{tabular}{lcc}
\hline\hline
\multicolumn{1}{l}{} &
\multicolumn{1}{c}{\wJ1311a} &
\multicolumn{1}{c}{\xJ1311b} \\
\multicolumn{1}{l}{Parameter} &
\multicolumn{1}{c}{Value and Error} &
\multicolumn{1}{c}{Value and Error} \\
\hline
$N_{\rm H, Gal}$ (10$^{20}$ cm$^{-2})$ & 4.95 (fixed)                             & 4.95 (fixed)                           \\
$N_{\rm H}$ (10$^{20}$ cm$^{-2})$      & --                                       & $45_{-29}^{+36}$                 \\
$\Gamma$                               & $1.3 \pm 0.4$                            & $1.6 \pm 0.5$                   \\
c-stat/d.o.f.                          & 201/510                                  & 297/509                                  \\
\hline
$F_{\rm 0.5-8~keV}$: obs, unabs & $9.9_{-2.7}^{+2.9}$, 10.3 & $10.7_{-3.1}^{+1.5}$, 14.5 \\
$F_{\rm 0.5-2~keV}$: obs, unabs & $2.4_{-0.8}^{+0.6}$, 2.7  & $2.0_{-0.8}^{+0.4}$, 5.3   \\
$F_{\rm 2-8~keV}$: obs, unabs   & $7.5_{-2.3}^{+3.0}$, 7.6  & $8.7_{-3.0}^{+2.2}$, 9.1   \\
\hline
\hline
\end{tabular}
\end{center}
\Chandra\ spectral parameters for the brightest X-ray source
within (\wJ1311a) and just outside and to the south of (\xJ1311b) the
2FGL error ellipse of \aJ1311. These are the \Suzaku\ detected sources,
named src A and src B, respectively, by \citet{mae11}. Fluxes ($F$) as
observed (obs) and unabsorbed (unabs) are in units of $10^{-14}$
\cgsflux. Errors quoted are at 90$\%$ confidence.
\label{table-j1311}
\end{table*}

\subsubsection{\aJ1311\ Field\label{sec-j1311}}

{\it CXOU~J131143.7-342749 (N31):} We found that this X-ray source is 
the only other source (the other case being \wJ1311a\ below) within the 
\aJ1311\ 2FGL ellipse with an optical counterpart. With only $\sim 4$ 
net counts detected in the $0.5-8$ keV band ($1.2 \sigma$), its spectrum 
is undersampled. The X-ray centroid, however, is well-determined ($r = 
0.26\arcsec$, statistical) and is offset by only 0.37\arcsec\ from an 
optical source (USNO~B1.0 0555-0290806; $B2 = 19.67$, $R2 = 18.87$, $I = 
18.33$ mag) which has a mid-IR counterpart (WISEP~J131143.72$-$342749.9; 
$W1 = 15.790$, $W2 = 15.302$, $W3 = 11.432$, $W4 = 7.954$ mag), but no 
near-IR counterpart in the 2MASS.

\begin{figure*}
  \begin{center}
    \includegraphics[width=6.25cm,angle=-90]{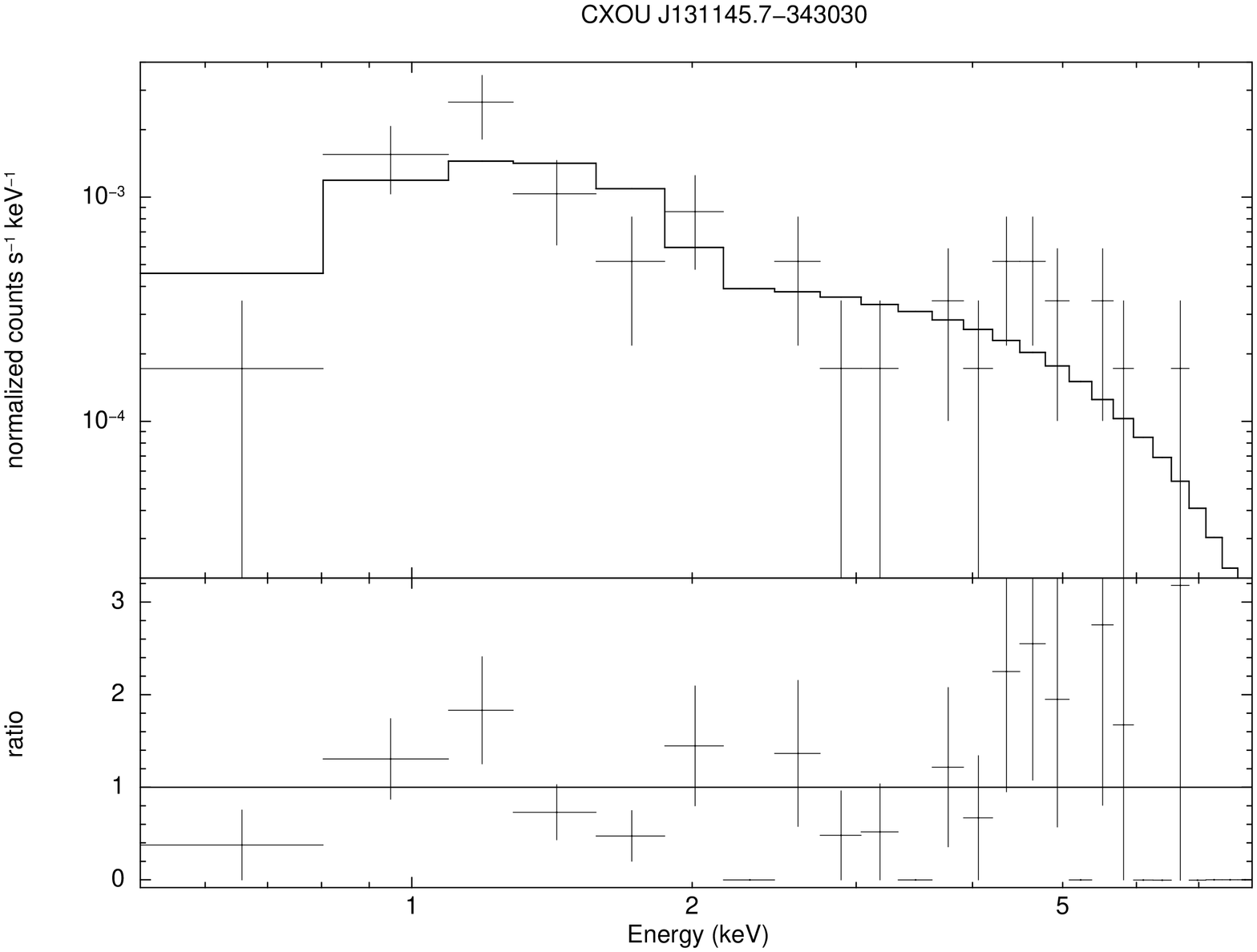}\includegraphics[width=6.25cm,angle=-90]{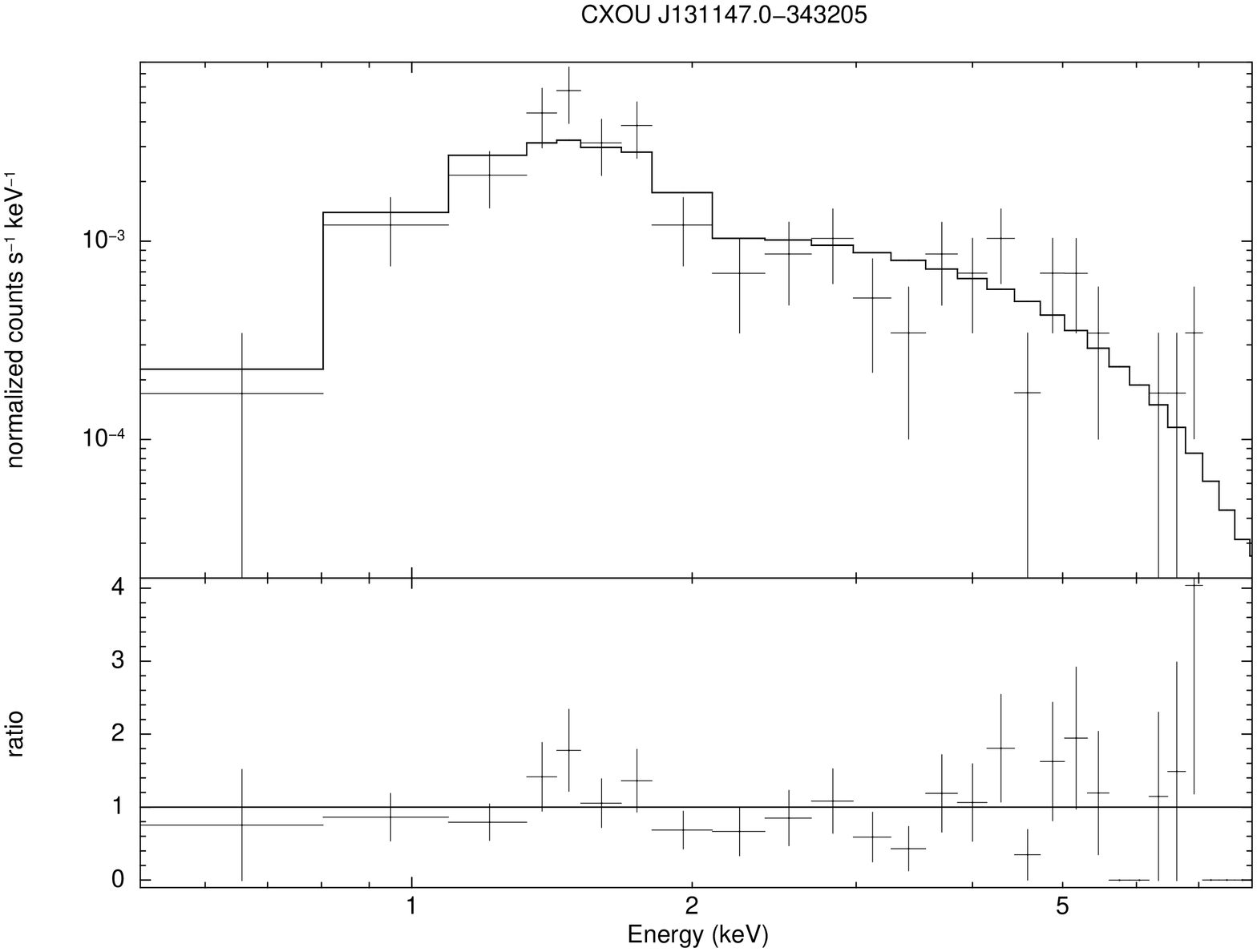}
  \end{center}
\caption{\Chandra\ ACIS-I spectrum of the brightest X-ray source within
the 2FGL error ellipse of \aJ1311 (\wJ1311a, N35; left), and of
the bright X-ray source just outside (to the south of) the error ellipse
(\xJ1311b, N39; right). The top panels show the data points with
the line indicating the best fit absorbed single power-law models and the
bottom panels show the ratio between the data and the models.
}
\label{spectrum1}
\end{figure*}

{\it \wJ1311a (N35):} This is the brightest X-ray source within the 2FGL 
error ellipse of \aJ1311. Its X-ray spectrum (Fig.~\ref{spectrum1}, 
left), is best fit with an absorbed single power-law with $\Gamma 
=1.3 \pm 0.4$ ($N_{\rm H, Gal} = 4.95 \times 10^{20}$ 
cm$^{-2}$), and an observed $0.5-8$ keV flux of $(9.9^{+2.9}_{-2.7}) 
\times 10^{-14}$ \cgsflux. Due to the low statistics and the already low 
$N_{\rm H, Gal}$ value, we found that the spectrum could also be fit 
without an absorption component, with $\Gamma = 1.1 \pm 0.4$. We quote 
the model parameters with the absorption included 
(Table~\ref{table-j1311}) to compare to the \Suzaku\ results from 
\citet{mae11}, who detected this source (named ``src A'' therein) on 
2009 Aug 04, seven months prior to our \Chandra\ observation. The 
\Suzaku\ data revealed short term variability in the form of a factor 
$\sim 10$ X-ray flare in the first 20 ks of the 100 ks duration 
observation (33 ks net exposure). The \Suzaku\ measured photon index 
($\Gamma = 1.38 \pm 0.13$) is consistent with that measured in our 
\Chandra\ observation, but the overall average $2-8$ keV flux of $(14.5 
\pm 1.8) \times 10^{-14}$ \cgsflux\ is $\sim 2\times$ larger. With the 
improved \Chandra\ point spread function over the \Suzaku\ one, we 
detect several additional fainter X-ray sources near this bright source 
(e.g., N38 and N33; Figure~\ref{image1}), but due to their low count 
rates, they can not be wholly responsible for the higher measured flux 
in the \Suzaku\ observation. \wJ1311a\ was also detected toward the edge 
of the FOV of a \Swift\ \citep{geh04} observation (3.34 ks, obs ID 
31358) from 2009 Feb 27 with the X-ray telescope \citep[XRT;][]{bur05}. 
The $0.3-10$ keV XRT count rate of $(4.5 \pm 1.4) \times 10^{-3}$ counts 
s$^{-1}$ is equivalent to a $0.5-8$ keV flux of $\sim 2 \times 10^{-13}$ 
\cgsflux, indicating a $\sim 2\times$ brighter source about one year 
prior to the \Chandra\ observation, providing further evidence that the 
X-ray source faded since 2009. We found that this bright \Chandra\ 
source has an optical counterpart (USNO~B1.0~0554-0289419, 0.62\arcsec\ 
offset) with $B2 = 21.02$ mag.

{\it \xJ1311b (N39):} Although this relatively bright X-ray source is 
outside (to the south) of the 2FGL error ellipse, we provide notes on it 
because it was the only other X-ray source detected in the \Suzaku\ 
observation mentioned above \citep[][``src B'' therein]{mae11}. For an 
absorbed single power-law fit to the \Chandra\ X-ray spectrum 
(Fig.~\ref{spectrum1}, right), we required an absorption component with 
a value, $N_{\rm H} = (45^{+36}_{-29}) \times 10^{20}$ cm$^{-2}$, well 
above the Galactic one. The \Chandra\ observed photon index and $2-8$ 
keV flux (Table~\ref{table-j1311}) are consistent with the \Suzaku\ 
measured values of $\Gamma = 1.34^{+0.16}_{-0.15}$ and 
$(12.0^{+1.8}_{-1.7}) \times 10^{-14}$ \cgsflux, indicating a relatively 
stable X-ray flux on months timescale. This source was outside the FOV 
of the \Swift\ XRT observation discussed above (obs ID 31358). With the 
improved \Chandra\ localization, we found a mid-infrared counterpart 
(WISEP~J131147.09$-$343205.3; $W1 = 15.720$, $W2 = 14.949$, $W3 = 
12.226$, $W4 = 9.056$ mag), but no corresponding source in the USNO~B1.0 
or 2MASS catalogs. Other than being relatively bright in X-rays, its 
location outside of the 2FGL error ellipse does not make it a 
particularly likely counterpart to the $\gamma$-ray source.

{\it CXOU~J131147.0$-$342953 (N40):} This faint X-ray source is the 
closest one detected to the 2FGL centroid (Figure~\ref{image1}). It has 
a possible mid-IR counterpart (WISEP~J131147.15-342953.5; $W1 = 16.846$, 
$W2 = 16.408$, $W3 = 12.443$, $W4 = 8.721$ mag) offset by 1.25\arcsec\ 
from the \Chandra\ position. We found no optical and near-IR counterpart 
to this source to the limit of the USNO~B1.0 and 2MASS catalogs, 
respectively.

The remaining detected X-ray sources within the 2FGL error ellipse have 
a range of $\sim 4-8$ net counts ($0.5-8$ keV), insufficient for 
detailed spectral or temporal analysis. None have optical/IR 
counterparts found within 2\arcsec\ of the \Chandra\ positions.  Of 
these, it is worth noting that the two most significant sources appear 
to have the hardest X-ray spectra, with the ratio of $2-8$ keV/$0.5-8$ 
keV counts, $\sim 6/7$ (CXOU~J131141.9$-$342955, N29) and $\sim 6/8$ 
(CXOU~J131146.4$-$343050, N38).

\begin{table*}
\begin{center}
\caption{}
\begin{tabular}{lcc}
\hline\hline
\multicolumn{1}{l}{} &
\multicolumn{1}{c}{\yJ1653a} &
\multicolumn{1}{c}{\zJ1653b} \\
\multicolumn{1}{l}{Parameter} &
\multicolumn{1}{c}{Value and Error} &
\multicolumn{1}{c}{Value and Error} \\
\hline
$N_{\rm H, Gal}$ (10$^{20}$ cm$^{-2})$ & 8.18 (fixed)                       & --      \\
$N_{\rm H}$ (10$^{20}$ cm$^{-2})$      & $9_{-9}^{+13}$               & $0_{-0}^{+27}$    \\
$\Gamma$                             & $1.8 \pm 0.3$               & --    \\
$kT$ (keV)                           & --       & $0.4^{+0.2}_{-0.1}$ \\
c-stat/d.o.f.                        & 335/509                              & 80/235    \\
\hline
$F_{\rm 0.5-8~keV}$: obs, unabs & $19.3_{-2.5}^{+2.1}$, 23.1 & $0.7_{-0.3}^{+0.2}$, -- \\
$F_{\rm 0.5-2~keV}$: obs, unabs & $6.0_{-1.2}^{+0.8}$,  9.5  & $0.5_{-0.2}^{+0.1}$, -- \\
$F_{\rm 2-8~keV}$: obs, unabs   & $13.3_{-2.2}^{+2.3}$, 13.6 & $0.2_{-0.2}^{+0.2}$, -- \\
\hline
\hline
\end{tabular}
\end{center}
As in Table~\ref{table-j1311}, but for the brightest (\yJ1653a) 
and second brightest (\zJ1653b) X-ray sources within the 2FGL error 
ellipse of \bJ1653.
\label{table-j1653}
\end{table*}

\subsubsection{\bJ1653\ Field\label{sec-j1653}}

{\it CXOU~J165315.6$-$015822 (N8):} This is the brightest X-ray source 
detected in the ACIS-I field of \bJ1653, but is $5.4'$ offset from the 
2FGL centroid and outside of the 95$\%$ confidence error ellipse ($r 
\simeq 3.6'$; see Fig.~\ref{image2}). It was probably detected 
previously as the faint RASS source 1RXS~J165312.9$-$015819 
\citep{vog00}, being only 41\arcsec\ away. The \Chandra\ position is 
coincident (0.50\arcsec\ offset) with the optical source 
USNO~B1.0~0880-0368721 ($B2 = 18.47$, $R2 = 16.97$, $I = 16.96$ mag), 
which is also detected in the near-IR (2MASS~J16531562$-$0158224; $J = 
16.196$, $H =15.560$, $K = 14.292$ mag) and mid-IR 
(WISEP~J165315.62$-$015822.3; $W1 = 12.599$, $W2 = 11.578$, $W3 = 
9.208$, $W4 = 7.364$ mag). Its X-ray spectrum (not shown) is well fit 
(c-stat/d.o.f.~= 427/510) with an absorbed power-law with $\Gamma = 1.86 
\pm 0.08$ ($N_{\rm H, Gal} = 8.18 \times 10^{20}$ cm$^{-2}$ fixed), and 
an observed $0.5-8$ keV flux $(14.1 \pm 0.6) \times 10^{-13}$ \cgsflux\ 
($15.7 \times 10^{-13}$ \cgsflux, unabsorbed). Leaving the absorption as 
a free parameter, we found negligible change in the photon index ($1.83 
\pm 0.13$), and the derived $N_{\rm H} = (7 \pm 5) \times 10^{20}$ 
cm$^{-2}$ converges toward the Galactic value. The absence of a radio 
counterpart in the NVSS ($\sim$2.5 mJy limit at 1.4 GHz) makes it 
unlikely to be a $\gamma$-ray emitting AGN. Although this source is 
quite prominent in X-rays, its placement outside the 2FGL 95$\%$ 
confidence ellipse means it is probably unrelated to \bJ1653.

{\it CXOU~J165337.2$-$020020 (N37):} This faint X-ray source near the 
measured 2FGL centroid is the only one within the LAT localizations of 
either source studied that is possibly variable ($P_{\rm KS} = 0.012$; 
Sec.~\ref{sec-fields}). Note however that this result is based on only 
$\sim 12$ net counts detected. There is no optical counterpart to the 
$\sim 21$ mag limit of the USNO~B1.0 catalog, nor was there a positional 
match in the 2MASS and WISEP catalogs.

\begin{figure*}
  \begin{center}
    \includegraphics[width=6.25cm,angle=-90]{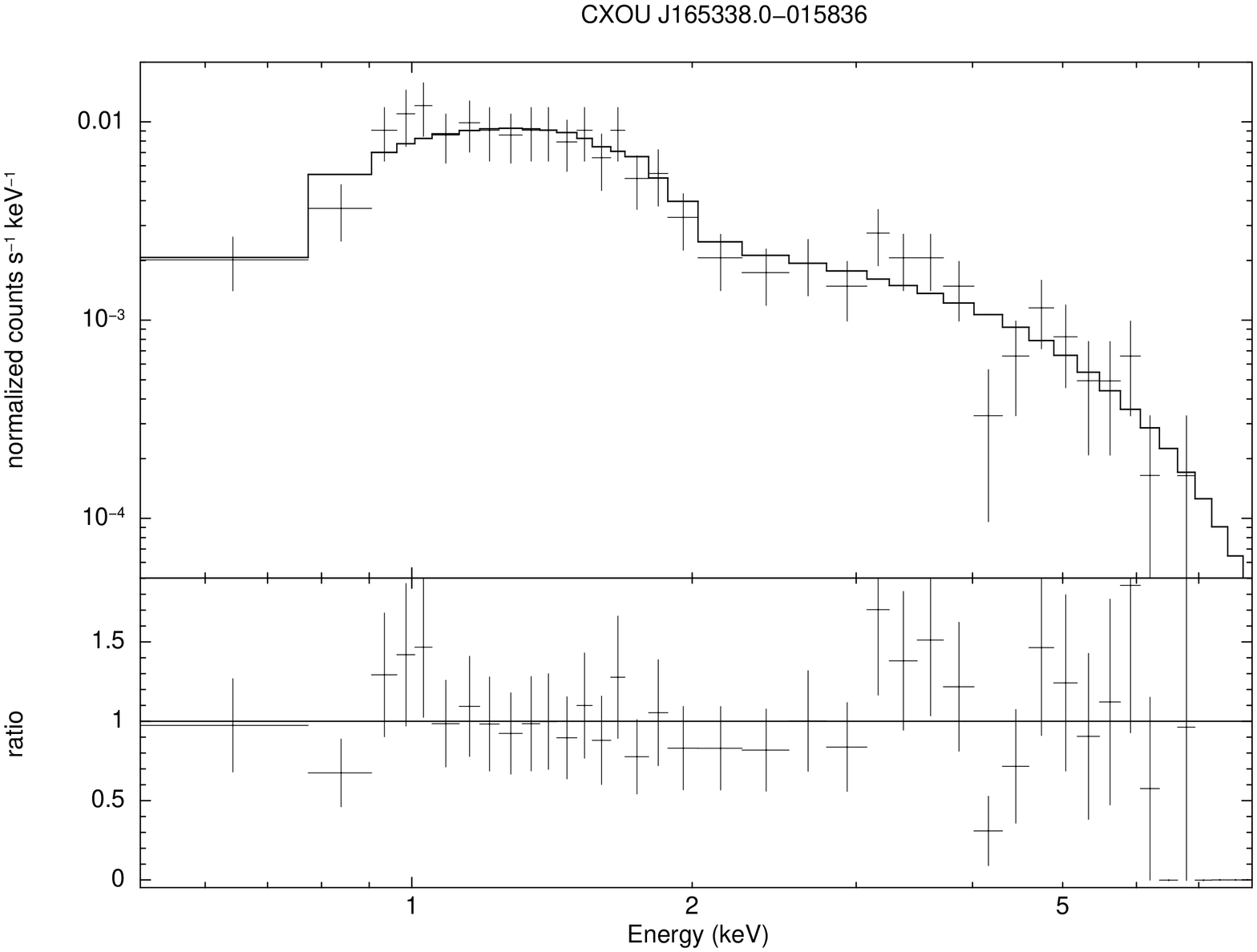}\includegraphics[width=6.25cm,angle=-90]{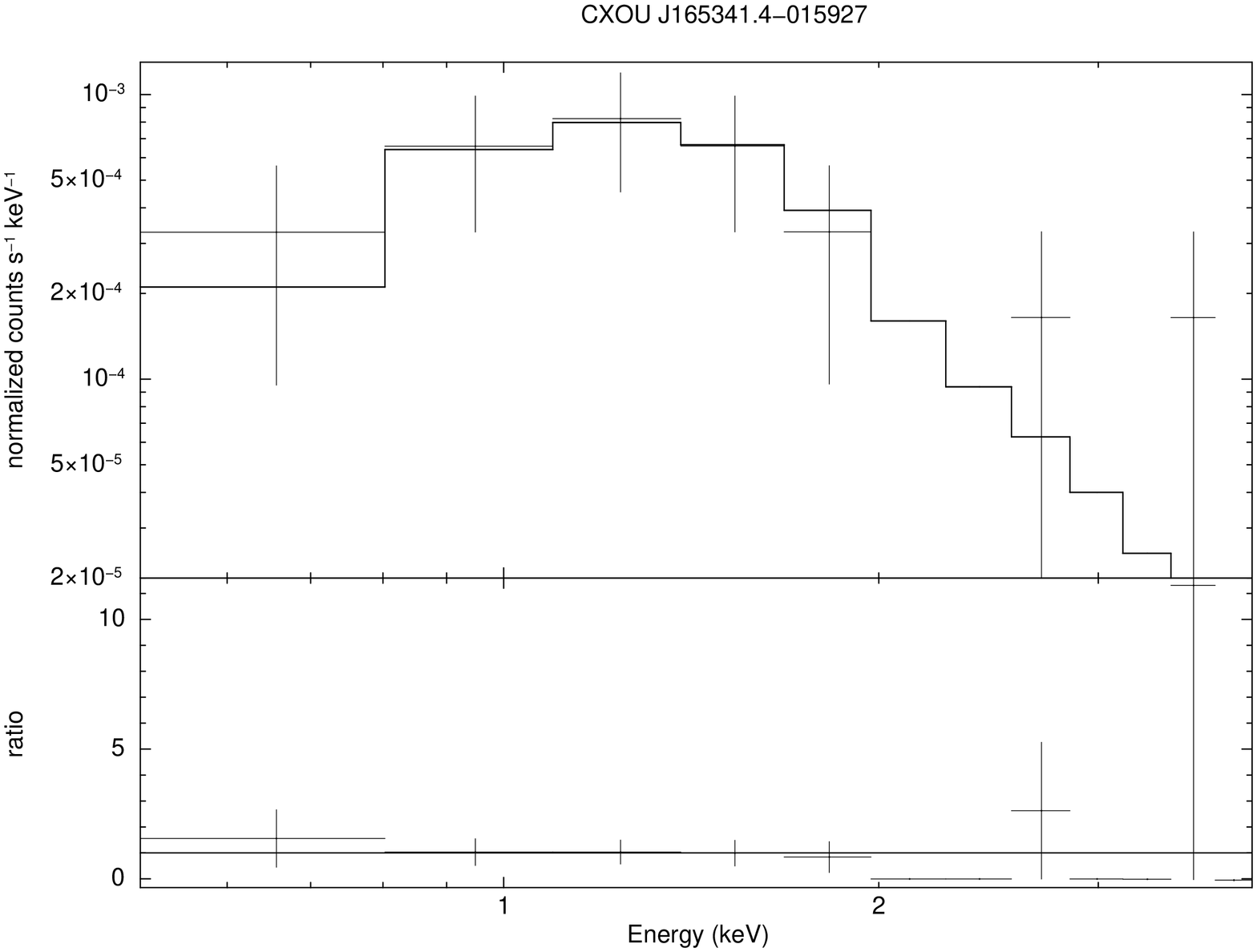}
  \end{center}
\caption{\Chandra\ ACIS-I spectrum of the brightest (\yJ1653a,
N38; left) and the second brightest (\zJ1653b, N44;
right) X-ray sources within the 2FGL error ellipse of \bJ1653. The top
panels show the data points with the line indicating the best fit
absorbed single power-law and blackbody models, respectively, and the
bottom panels show the ratio between the data and the models.
}
\label{spectrum2}
\end{figure*}

{\it \yJ1653a (N38):} This is the brightest X-ray source found within 
the 2FGL error ellipse of \bJ1653. To its X-ray spectrum 
(Fig.~\ref{spectrum2}, left), we fitted an absorbed single power-law 
with $\Gamma = 1.8 \pm 0.3$, and found a hint of an additional 
absorption component with $N_{\rm H} = (9^{+13}_{-9}) \times 
10^{20}$ cm$^{-2}$ larger than the Galactic value ($N_{\rm H, Gal} = 
8.18 \times 10^{20}$ cm$^{-2}$). We found that a pure power-law model 
without any absorption overestimates the soft part of the spectrum, thus 
was a poor model for the data. This X-ray source was also detected in a 
snapshot \Swift\ XRT observation (4.85 ks, obs ID 31379) from 2009 Mar 
22 with a $0.3-10$ keV count rate of $(3.2 \pm 1.1) \times 10^{-3}$ 
counts s$^{-1}$. This is equivalent to a $0.5-8$ keV flux of $\sim 16 
\times 10^{-14}$ \cgsflux, indicating a roughly stable flux relative to 
the \Chandra\ observed value, $(19.3^{+2.1}_{-2.5}) \times 10^{-14}$ 
\cgsflux\ (Table~\ref{table-j1653}) obtained $\sim 10$ months later. 
This X-ray source has an optical counterpart, USNO-B1.0~0880-0369025 
($B2 = 20.40$, $R2 = 19.41$, and $I = 20.0$ mag), which is only 
0.28\arcsec\ offset from the \Chandra\ determined centroid.

{\it \zJ1653b (N44):} This is the second brightest X-ray source found 
within the 2FGL error ellipse of \bJ1653. Only $\sim 19$ net counts 
($0.5-8$ keV) were found, with most of the counts (17) in the $0.5-2$ 
keV range indicating a soft X-ray spectrum. Despite the low statistics, 
the X-ray spectrum can be fit with a blackbody model 
(Fig.~\ref{spectrum2}, right), with temperature, $kT = 
0.4^{+0.2}_{-0.1}$ keV, and absorption consistent with zero 
(Table~\ref{table-j1653}). We found an optical source, 
USNO-B1.0~0880-0369077 ($B2 = 20.58$, $R2 = 18.95$, and $I = 17.05$ mag) 
0.68\arcsec\ offset from the X-ray centroid. The USNO source has a 
near-infrared (2MASS~J16534140$-$0159272; $J = 15.094$, $H = 14.469$, $K 
= 14.147$ mag) and a mid-infrared (WISEP~J165341.41$-$015927.6; $W1 = 
14.057$, $W2 = 13.843$, $W3 = 12.569$, $W4 = 8.667$ mag) counterpart.

{\it CXOU~J165342.8$-$020144 (N49):} This is the only other X-ray source 
within the 2FGL ellipse of \bJ1653\ (other than the two mentioned 
sources above, N38 and 44) with an USNO~B1.0 catalog match. The X-ray 
source is offset by 0.32\arcsec\ from the relatively bright optical 
source ($B2 = 16.64$, $R2 = 14.60$, $I = 14.16$ mag), 
USNO~B1.0~0879-0416658. The USNO source has a near-infrared 
(2MASS~J16534285$-$0201450; $J = 13.133$, $H = 12.575$, $K = 12.492$ 
mag) and mid-infrared (WISEP~J165342.83$-$020145.1; $W1 = 12.420$, $W2 = 
12.429$, $W3 = 12.292$, $W4 = 9.014$ mag) counterpart. Although the 
source is faint ($1.6\sigma$ in $0.5-8$ keV band), all six of its 
detected counts are in the $0.5-2$ keV range indicating a soft spectrum.

{\it CXOU~J165343.4$-$015841 (N51):} The X-ray spectrum of this faint 
source appears soft with all of its $\sim 6$ detected net counts in the 
$0.5-2$ keV band. It is the only case of a detected X-ray source within 
the 2FGL ellipse of \bJ1653\ that has a mid-IR counterpart (0.7\arcsec\ 
offset from WISEP~J165343.52$-$015840.6; $W1 = 16.245$, $W2 = 16.366$, 
$W3 = 12.612$, $W4 = 8.882$ mag) and without a match in the USNO~B1.0 
and 2MASS catalogs.

Amongst the remaining detected X-ray sources within the 2FGL error 
ellipse, only $\sim 5-14$ net counts ($0.5-8$ keV) are observed per 
object. Although the statistics are limited, we found that several 
sources have most of their detected counts in either the $0.5-2$ keV or 
$2-8$ keV range, indicating soft and hard spectra, respectively. As was 
the case for the two faint sources noted above (CXOU~J165342.8$-$020144 
= N49 and CXOU~J165343.4$-$015841 = N51), we found that 
CXOU~J165328.4$-$020009 (N24) has an indication for a soft spectrum, 
with a ratio of $0.5-2$ keV/$0.5-8$ keV counts, $\sim 12/14$. Similarly, 
the hardest spectrum source has a ratio of $2-8$ keV/$0.5-8$ keV counts, 
$\sim 5/6$ (CXOU~J165349.4$-$015818 = N64). The latter two sources do 
not have optical/IR counterparts at the sensitivity limits of the 
USNO~B1.0, 2MASS, and WISEP catalogs.

\section{Discussion\label{sec-discussion}}

The $\gamma$-ray sources studied, \aJ1311\ and \bJ1653, are the two 
brightest unidentified \Fermi-LAT sources found at high Galactic 
latitudes.  In fact, both have been detected by EGRET 
(Sec.~1), and the improved localizations provided now by 
the \Fermi-LAT with \r95 $\simeq~2.0' - 3.6'$ \citep[compared to the 
corresponding EGRET values $\simeq~34' - 44'$;][]{har99} allow us to 
address the counterparts with more certainty. From an X-ray perspective, 
our \Chandra\ observations detected sources down to a $0.5-8$ keV flux 
threshold of $\sim (0.2-0.3) \times 10^{-14}$ \cgsflux\ 
(Table~\ref{table-2}). This is much improved over typical flux limits of 
$\sim 10^{-13}$ \cgsflux\ achieved in all-sky surveys like the RASS 
\citep[0.1$-$2.4 keV;][]{vog99,vog00}. Also, existing pointed \Swift\ 
and \Suzaku\ observations (Sec.~\ref{sec-bright}) detected only the 
single brightest source within the LAT error ellipses, compared to the 9 
and 13 \Chandra\ detected X-ray sources that can now be considered as 
potential counterparts to the LAT $\gamma$-ray source.

Due to their high Galactic latitudes, the most obvious candidate 
counterparts to these unidentified $\gamma$-ray sources would be 
extragalactic objects, specifically, blazars fainter than currently 
catalogued. The LAT error circles of the high latitude objects have in 
fact been searched for blazars down to a radio flux limit of $\sim$30 
mJy \citep{lbas}, but even fainter blazars are now being found in large 
numbers, e.g., from cross-correlations of the SDSS/RASS/FIRST databases 
\citep[e.g.,][]{plo08}. In this context however, we found that none of 
the detected X-ray sources in either field had radio counterparts in the 
NVSS catalog (Sec.~\ref{sec-bright}). In fact, only one (extended) radio 
source was found within the 95$\%$ LAT error ellipse of \bJ1653\ 
(below), with none detected within the \aJ1311\ localization. The NVSS 
flux limit of $\sim$2.5 mJy at 1.4 GHz is $\sim 10 \times$ fainter than 
the faintest typical radio sources currently associated with \Fermi-LAT 
$\gamma$-ray blazars \citep{2lac,radgamma}. This absence of radio 
sources in general, and the lack of point source counterparts to the 
X-ray detected objects specifically, allow us to rule out faint blazars 
as possible counterparts of these sources. Radio-quiet AGN are not 
currently known to be $\gamma$-ray emitters, except for a few examples 
of nearby galaxies with substantial starburst contributions 
\citep{2lac,len10,ten11,latseyferts}, so even if the \Chandra\ detected 
X-ray sources are confirmed to be AGN, they are likely unrelated to the 
$\gamma$-ray source.

From a $\gamma$-ray perspective, the faintest radio blazars tend to have 
harder HE $\gamma$-ray spectra ($\Gamma <2$), sometimes extending into 
TeV energies. Unlike these blazars, the two $\gamma$-ray sources 
discussed are characterized by soft $0.1-100$ GeV spectra, being 
parameterized with single power-laws in the 1FGL catalog analysis 
\citep{1fgl} with slopes, $\Gamma = 2.25 \pm 0.05$ (\aJ1311) and $2.29 
\pm 0.06$ (\bJ1653). In fact, a more detailed analysis of the longer 2 
year LAT dataset presented in the 2FGL catalog indicated the HE spectra 
were best characterized as log-parabolas 
\citep{2fgl}\footnote{http://heasarc.gsfc.nasa.gov/FTP/fermi/data/lat/catalogs/ source/lightcurves/2FGL$\_$J1311d7m3429$\_$spec.png 
http://heasarc.gsfc.nasa.gov/FTP/fermi/data/lat/catalogs/ source/lightcurves/2FGL$\_$J1653d6m0159$\_$spec.png 
\label{footnote-spec}}, not typically observed in $\gamma$-ray blazars 
\citep{lbasspectra}. Moreover, unlike typical bright $\gamma$-ray 
emitting blazars, their variability indices of $17-19$ are below the 
41.6 threshold in the 2FGL catalog analysis \citep{2fgl}, indicating no 
significant $\gamma$-ray variability within the 24 month 
dataset\footnote{http://heasarc.gsfc.nasa.gov/FTP/fermi/data/lat/catalogs/ source/lightcurves/2FGL$\_$J1311d7m3429$\_$lc.png 
http://heasarc.gsfc.nasa.gov/FTP/fermi/data/lat/catalogs/ source/lightcurves/2FGL$\_$J1653d6m0159$\_$lc.png 
\label{footnote-lc}}.

Nearby radio galaxies have been found to be likely $\gamma$-ray emitters 
\citep[e.g.,][and references therein]{2lac} and it is possible that some 
fraction of the unidentified high Galactic latitude LAT sources could be 
unknown radio galaxies (i.e., `misaligned blazars') that could be faint 
X-ray emitters \citep[e.g.,][]{can99}. In particular, steady 
$\gamma$-ray emission from radio lobes is possible \citep{che07}, as was 
observed in the nearby radio galaxy Centaurus~A \citep{cena} and 
possibly in NGC~6251 \citep{tak12b}. In this context, the single radio 
source (NVSS~J165348.44$-$015958.7; 11.5 $\pm$ 1.7 mJy at 1.4 GHz) found 
within the \bJ1653\ error ellipse is extended, with measured elliptical 
dimensions, major axis = 125.4\arcsec, minor axis $<46.9\arcsec$ (i.e., 
unresolved in this direction), at position angle = 36\deg. Inspecting 
Figure~\ref{image2}, we see that the \Chandra\ sources N58 and N57 
(within the 2FGL ellipse) and N69 (outside the 2FGL ellipse) are located 
near the extended tips of this radio source. If any of these \Chandra\ 
sources are related to NVSS~J165348.44$-$015958.7, it could plausibly 
mark a radio outflow from the X-ray source. However, the lack of an 
optically bright extended giant elliptical galaxy counterpart to any of 
these X-ray sources eliminates the possibility that these are nearby 
radio galaxies. Relatedly, young radio sources are expected to be steady 
$\gamma$-ray emitters \citep[e.g.,][]{mcc11}, but such objects are 
typically bright compact cm-wavelength sources but no such sources were 
found in the NVSS image within the LAT error regions in the present 
cases.

As AGN likely can be ruled out as potential counterparts, the remaining 
possibilities are open to debate. As both $\gamma$-ray sources are 
located at mid Galactic latitudes ($|b| =10\deg - 30\deg$; 
Table~\ref{table-1}), there is the interesting possibility of isolated 
neutron stars associated with the $\gamma$-ray sources, as was posited 
for 3EG~J1835+5918 \citep[][and references therein]{hal02}, and 
subsequently confirmed with \Fermi-LAT observations of this 
\citep{j1836} and other pulsars \citep{psrcat}. Indeed, \aJ1311\ and 
\bJ1653\ are two of nine total high Galactic latitude ($|b|>10\deg$) 
$\gamma$-ray sources from the LAT bright source list \citep[][]{bsl} 
that were unidentified at the time. The other seven sources have all 
since been identified as pulsar powered sources. Specifically, they were 
predominantly identified as radio/$\gamma$-ray emitting MSPs -- 
PSR~J0614$-$3329, PSR~J1231$-$1411, PSR~J2214+3000 \citep{ran11}, 
PSR~J2241$-$5236 \citep{kei11}, and PSR~J2302+4442 \citep{cog11} -- with 
one normal young pulsar \citep[PSR~J2055+25;][]{saz10}. The two subjects 
of the present study have been similarly searched for pulsating radio 
emission with null results in the past \citep[based on their EGRET 
localizations;][]{cra06} and in new searches of the \Fermi\ error 
circles \citep{ran11}. In the remaining case (\eJ2339), the brightest 
\Chandra\ X-ray source within the \Fermi-LAT error ellipse was found to 
be a black widow-type MSP and is the likely counterpart of the 
$\gamma$-ray source \citep[][see also \citet{rom11}]{kon12}. As in these 
other seven 0FGL cases, the LAT spectra of our remaining two 
unidentified objects display significant 
curvature$^{\ref{footnote-spec}}$ and are steady $\gamma$-ray 
emitters$^{\ref{footnote-lc}}$, so may point to a pulsar origin for them 
as well.

In our \Chandra\ observations of the MSPs identified with the LAT 
sources, PSR~J2214+3000 \citep{ran11} and PSR J2241$-$5236 
\citep{kei11}, the MSPs were spatially coincident with the brightest 
X-ray sources within the LAT error ellipses. The X-ray spectral analysis 
indicated a thermal origin with best fit blackbody temperatures, $kT 
\sim 0.2-0.3$ keV, with essentially no photons above 2 keV. Similar 
X-ray spectral results for the other identified pulsars were derived 
from \XMM\ observations of PSR~J2302+4442 \citep{cog11} and \Swift\ 
observations of PSR~J0614$-$3329 and PSR~J1231$-$1411 \citep{ran11}; see 
also past \XMM\ observations of the 0FGL~J0614.3$-$3330 case 
\citep{lap06}. In fact, we found that several X-ray sources 
(Sec.~\ref{sec-bright}) have soft spectra that appear thermal in origin. 
Taking the 2FGL \citep{2fgl} observed $0.1-100$ GeV $\gamma$-ray fluxes 
for \aJ1311\ ($F_{\gamma} = 6.17 \times 10^{-11}$ \cgsflux) and \bJ1653\ 
($F_{\gamma} = 3.43 \times 10^{-11}$ \cgsflux), the \Chandra\ 
observations probe a range of X-ray\footnote{To ease comparison with 
other works \citep[see, e.g.,][]{mar11}, we extrapolated our $0.5-8$ keV 
X-ray fluxes ($F_{\rm 0.5-8~keV}$) into $0.3-10$ keV X-ray ones ($F_{\rm 
X}$) assuming, $F_{\rm X} = 1.25 \times F_{\rm 0.5-8~keV}$.} to 
$\gamma$-ray flux (or luminosity) ratios, $F_{\rm X}/F_{\gamma} \simlt 
(2-8) \times 10^{-3}$ corresponding to the brightest X-ray sources, and 
down to $\sim (0.04 - 0.1) \times 10^{-3}$ for the faintest ones. For a 
pulsar interpretation, this probes the conversion ranges of spin-down 
powers to (non-thermal or thermal) X-rays for possible neutron star 
candidates \citep[e.g.,][]{mir00}.

In contrast to the thermal X-ray sources discussed above, the brightest 
\Chandra\ sources in our two cases show X-ray spectra that extend to 
$\sim 7-8$ keV and were best fit with single power-law spectra 
(Sec.~\ref{sec-bright}). If these are the counterparts to the LAT 
sources, the objects are likely not normal pulsars or MSPs. The 
non-thermal X-ray spectra of these sources are similar to the case of 
CXOU~J233938.7$-$053305, the putative counterpart of the (formerly) 
unidentified high Galactic latitude source \eJ2339\ \citep{kon12}. In 
the latter case, the spectrum is best characterized by an absorbed 
power-law model with $\Gamma = 1.1$ and a $0.3-10$ keV X-ray flux of $3 
\times 10^{-13}$ \cgsflux. Its black widow-like MSP nature was 
discovered after optical follow-up of the detected X-ray variable 
\Chandra\ source revealed a 4.63 hr period of the binary derived via 
optical photometric \citep{kon12} and spectroscopic \citep{rom11} 
monitoring. In our \Chandra\ observations, we found only one faint X-ray 
source (CXOU~J165337.2$-$020020, N37) within the \bJ1653\ error ellipse 
to be possibly variable, albeit with limited statistics. In the case of 
\wJ1311a, the bright source within the 2FGL ellipse of \aJ1311\ that had 
a \Suzaku\ discovered short term flare (factor of 10 increase in the the 
first 20 ks of the $\sim$100 ks observation span), we found no 
significant variability within our 20 ks \Chandra\ observation. However, 
these observations together with a \Swift\ snapshot, provided evidence 
for variability on months timescales also (Sec.~\ref{sec-bright}). In 
this context, it may be fruitful to photometrically monitor the optical 
counterparts to these X-ray sources (Sec.~\ref{sec-bright}) for similar 
variability as in the case of CXOU~J233938.7$-$053305.

\section{Summary\label{sec-summary}} 

\Chandra\ observations of the two brightest unidentified high Galactic 
latitude $\gamma$-ray sources from the 3 month \Fermi-LAT bright source 
list were obtained. Both sources were previously detected by EGRET, and 
remain two of the most enigmatic $\gamma$-ray sources to date. The basic 
X-ray properties of all sources observed in the ACIS-I FOV were 
determined, with 9 to 13 X-ray sources detected within the respective 
\Fermi-LAT error ellipses. Using existing catalogs (USNO~B1.0, 2MASS, 
WISE), the sub-arcsecond \Chandra\ positions enabled us to locate 
optical, near-infrared and mid-infrared counterparts to several of the 
X-ray sources. The ensemble of X-ray sources, focused primarily on the 
ones within the \Fermi-LAT localizations, were further discussed as 
potential counterparts of the $\gamma$-ray sources. Although existing 
all-sky optical/infrared catalogs returned only a handful of X-ray 
counterpart matches, our work enables future optical identifications if 
deeper images can be obtained.

In the case of \aJ1311, the brightest X-ray source (\wJ1311a) within the 
\Fermi-LAT error ellipse is the most credible counterpart. This source 
was detected in a previous \Suzaku\ observation with X-ray variability 
on sub-day timescales, and our newly presented \Chandra\ (and \Swift) 
observations suggest variability on longer timescales. Together with the 
power-law nature of the X-ray spectrum, it appears similar to the case 
of CXOU~J233938.7$-$053305, a candidate black widow MSP that is the 
likely counterpart of a similar \Fermi-LAT source, \eJ2339\ 
\citep{kon12}. Another bright X-ray source, \xJ1311b, found just outside 
the LAT error ellipse and also previously detected by \Suzaku, did not 
show any significant X-ray variability, and is a less probable 
counterpart to the $\gamma$-ray source.

In the case of \bJ1653, the brightest X-ray source (\yJ1653a) within the 
\Fermi-LAT localization also displays a non-thermal X-ray spectrum from 
our \Chandra\ observation. However, there was no significant X-ray 
variability found within our 20 ks observation, and on the longer 
timescale probed by comparing to a \Swift\ observation obtained $\sim$10 
months earlier. A faint optical counterpart to this X-ray source is 
found, and optical photometric and spectroscopic monitoring may be 
fruitful to test if this source is similar to the case of 
CXOU~J233938.7$-$053305. The second brightest \Chandra\ source 
(CXOU~J165341.4$-$015927) found within the 2FGL localization also has an 
optical/infrared counterpart, and has a likely thermal origin for the 
X-rays. This, and the other fainter X-ray sources revealed in our 
\Chandra\ observation, can be similarly investigated to ultimately probe 
the nature of this unidentified $\gamma$-ray source.

\acknowledgments

\begin{center}Acknowledgments\end{center}

We thank S.~Digel, M.~Wolff, E.~Hays, and \L.~Stawarz for comments on the 
manuscript, T.~Burnett, E.~Ferrara, E.~Hays, Y.~Kanai, J.~Kataoka, 
N.~Kawai, M.~Kerr, T.~Nakamori, P.~Ray, D.~J.~Thompson, N.~Vilchez, and 
M.~T.~Wolff for their support and participation in the \Chandra\ 
observing proposal, and members of the \Fermi-LAT team for providing the 
preliminary LAT localizations utilized over the course of this work.

This work began while C.~C.~C.~was supported by an appointment to the 
NASA Postdoctoral Program at Goddard Space Flight Center, administered 
by Oak Ridge Associated Universities through a contract with NASA; his 
work at NRL is supported in part by NASA DPR S-15633-Y. Support for this 
work was partially provided by the National Aeronautics and Space 
Administration through \Chandra\ Award Number GO0-11022A (C.~C.~C., D.~D.) 
issued by the \Chandra\ X-ray Observatory Center, which is operated by the 
Smithsonian Astrophysical Observatory for and on behalf of the National 
Aeronautics Space Administration under contract NAS8-03060.

{\it Facilities:} \facility{CXO ()}, \facility{Swift ()}

{}


\clearpage
\newpage

\appendix

\section{A. X-ray Source Lists}

This section lists the X-ray properties of all sources detected in the 
four ACIS-I chips of the \Chandra\ fields (see Sec.~\ref{sec-fields}). 
In Table~\ref{table-all1311} (\aJ1311) and Table~\ref{table-all1653} 
(\bJ1653), sources are listed with their catalog numbers (N) in order of 
increasing R.A. with coordinates in J2000.0. Listed also are the source 
localization errors ($r$), its distance from the \Chandra\ aim-point 
($D$), the effective area ($A_{\rm eff}$) at the source position at 1.5 
keV (lower values indicate sources located closer to the chip gaps and 
edges), the logarithm probability from the KS-test for variability (log 
$P_{\rm KS}$), and the net counts, significances ($\sigma$), and 
probability ($P_{\rm B}$) that the source counts are solely from the 
background in the $0.5-8$ keV, $0.5-2$ keV, and $2-8$ keV bands. 
Following \citet{bro10}, we considered only sources with more than four 
net counts in the full band and not near the CCD chip edges, and found 
five sources in each of the ACIS-I fields to be possibly variable 
($0.005 < P_{\rm KS} < 0.05$) -- \aJ1311\ (N34, N49, N76, N88, N89) and 
\bJ1653\ (N14, N37, N70, N81, N92). There is no evidence for variability 
($P_{\rm KS} > 0.05$) in the remaining sources. The analysis described 
in Sec.~\ref{sec-fields} also provides an afterglow fraction, which if 
$>$0, indicates that the source is possibly contaminated by afterglow 
events and their significance could be lower than quoted and the source 
position, spectrum, and/or lightcurve are potentially affected. The 
sources flagged in this way were:
J131145.71-343030.5 (N35),
J131147.03-343205.2 (N39),
J131226.81-343029.7 (N93),
J165315.60-015822.1 (N8),
J165323.22-020451.2 (N14),
J165325.88-015107.7 (N19),
J165333.54-015259.5 (N28),
J165338.05-015836.6 (N38), and
J165359.02-020316.0 (N75).

\begin{sidewaystable}[ht]
\footnotesize
\begin{center}
\caption{}
    \tabcolsep 5.0pt
\begin{tabular}{lcccccccccccccccc}
\hline\hline
\multicolumn{1}{c}{N} &
\multicolumn{1}{c}{CXOU J} &
\multicolumn{1}{c}{R.A.} &
\multicolumn{1}{c}{Decl.} &
\multicolumn{1}{c}{$r$} &
\multicolumn{1}{c}{$D$} &
\multicolumn{1}{c}{$A_{\rm eff}$} &
\multicolumn{1}{c}{log} &
\multicolumn{1}{c}{Counts} &
\multicolumn{1}{c}{$\sigma$} &
\multicolumn{1}{c}{$P_{\rm B}$} &
\multicolumn{1}{c}{Counts} &
\multicolumn{1}{c}{$\sigma$} &
\multicolumn{1}{c}{$P_{\rm B}$} &
\multicolumn{1}{c}{Counts} &
\multicolumn{1}{c}{$\sigma$} &
\multicolumn{1}{c}{$P_{\rm B}$} \\
\cline{9-11}\cline{12-14}\cline{15-17}
\multicolumn{1}{c}{} &
\multicolumn{1}{c}{Name} &
\multicolumn{1}{c}{(deg)} &
\multicolumn{1}{c}{(deg)} &
\multicolumn{1}{c}{($\arcsec$)} &
\multicolumn{1}{c}{($'$)} &
\multicolumn{1}{c}{(cm$^2$)} &
\multicolumn{1}{c}{$P_{\rm KS}$} &
\multicolumn{3}{c}{($0.5-8$ keV)} &
\multicolumn{3}{c}{($0.5-2$ keV)} &
\multicolumn{3}{c}{($2-8$ keV)} \\
\hline
  1 & 131105.8-342652 & 197.77440 & -34.44785 & 1.36 & 9.7 & 401 & -0.31 &   10.4 +5.0/-3.8 & 2.1 & 1.4e-04 &    6.3 +4.0/-2.8 & 1.6 & 5.3e-04 &    4.1 +3.8/-2.6 & 1.1 & 3.4e-02 \\ 
  2 & 131108.7-342738 & 197.78660 & -34.46072 & 0.93 & 8.9 & 414 & -0.12 &   17.9 +5.7/-4.6 & 3.1 & 9.3e-11 &    7.9 +4.1/-2.9 & 1.9 & 7.1e-06 &   10.0 +4.6/-3.4 & 2.2 & 2.5e-06 \\ 
  3 & 131111.7-342856 & 197.79906 & -34.48245 & 1.26 & 8.2 & 390 & -0.00 &    6.7 +4.1/-2.9 & 1.6 & 8.4e-04 &    4.3 +3.4/-2.2 & 1.2 & 1.3e-03 &    2.4 +3.2/-1.9 & 0.8 & 8.1e-02 \\ 
  4 & 131112.2-343022 & 197.80102 & -34.50632 & 0.65 & 8.2 & 437 & -0.16 &   26.5 +6.5/-5.4 & 4.1 & 6.0e-20 &   14.1 +5.0/-3.8 & 2.8 & 6.4e-13 &   12.4 +4.8/-3.7 & 2.6 & 5.3e-09 \\ 
  5 & 131116.4-343147 & 197.81843 & -34.52992 & 0.99 & 7.7 & 439 & -0.56 &    9.1 +4.4/-3.3 & 2.0 & 8.8e-06 &    6.2 +3.8/-2.6 & 1.6 & 4.6e-05 &    2.9 +3.2/-1.9 & 0.9 & 2.7e-02 \\ 
  6 & 131118.0-342952 & 197.82502 & -34.49800 & 0.54 & 6.9 & 447 & -0.02 &   19.0 +5.6/-4.4 & 3.4 & 1.1e-18 &   11.7 +4.6/-3.4 & 2.5 & 1.6e-14 &    7.3 +4.0/-2.8 & 1.9 & 1.0e-06 \\ 
  7 & 131118.1-343013 & 197.82542 & -34.50362 & 0.81 & 6.9 & 449 & -0.08 &    8.9 +4.3/-3.1 & 2.1 & 5.7e-07 &    3.6 +3.2/-1.9 & 1.1 & 6.3e-04 &    5.2 +3.6/-2.4 & 1.4 & 2.2e-04 \\ 
  8 & 131119.6-342935 & 197.83179 & -34.49324 & 1.21 & 6.6 & 454 & -0.48 &    3.1 +3.2/-1.9 & 1.0 & 1.5e-02 &   -0.3 +1.9/0.0 & -0.2 & 1.0 &    3.4 +3.2/-1.9 & 1.1 & 3.7e-03 \\ 
  9 & 131122.0-342744 & 197.84203 & -34.46232 & 1.07 & 6.2 & 388 & -0.78 &    3.1 +3.2/-1.9 & 1.0 & 1.3e-02 &    2.7 +2.9/-1.6 & 0.9 & 4.8e-03 &    0.4 +2.3/-0.8 & 0.2 & 4.4e-01 \\ 
 10 & 131125.5-342726 & 197.85662 & -34.45749 & 0.73 & 5.6 & 388 & -0.25 &    4.6 +3.4/-2.2 & 1.3 & 1.0e-04 &    0.8 +2.3/-0.8 & 0.4 & 1.5e-01 &    3.7 +3.2/-1.9 & 1.2 & 2.3e-04 \\ 
 11 & 131126.8-343428 & 197.86202 & -34.57448 & 1.40 & 7.3 & 455 & -0.76 &    3.3 +3.4/-2.2 & 1.0 & 3.3e-02 &   -0.5 +1.9/0.0 & -0.3 & 1.0 &    3.8 +3.4/-2.2 & 1.1 & 8.8e-03 \\ 
 12 & 131126.9-343350 & 197.86211 & -34.56416 & 0.75 & 6.9 & 461 & -0.88 &   11.6 +4.7/-3.6 & 2.5 & 4.8e-09 &    9.4 +4.3/-3.1 & 2.2 & 2.4e-09 &    2.2 +2.9/-1.6 & 0.8 & 4.5e-02 \\ 
 13 & 131129.4-343133 & 197.87262 & -34.52597 & 0.69 & 5.1 & 483 & -0.22 &    4.5 +3.4/-2.2 & 1.3 & 1.4e-04 &   -0.2 +1.9/0.0 & -0.1 & 1.0 &    4.7 +3.4/-2.2 & 1.4 & 2.2e-05 \\ 
 14 & 131131.4-343415 & 197.88124 & -34.57095 & 1.07 & 6.5 & 451 & -0.04 &    4.7 +3.6/-2.4 & 1.3 & 2.7e-03 &    4.6 +3.4/-2.2 & 1.4 & 6.9e-05 &    0.1 +2.3/-0.8 & 0.0 & 6.1e-01 \\ 
 15 & 131131.8-342546 & 197.88264 & -34.42969 & 0.78 & 5.3 & 457 & -0.03 &    3.4 +3.2/-1.9 & 1.1 & 3.2e-03 &    1.8 +2.7/-1.3 & 0.7 & 1.7e-02 &    1.6 +2.7/-1.3 & 0.6 & 6.2e-02 \\ 
 16 & 131133.7-342251 & 197.89081 & -34.38105 & 1.53 & 7.3 & 419 & -0.56 &    2.5 +3.4/-2.2 & 0.7 & 1.1e-01 &   -0.8 +1.9/0.0 & -0.4 & 1.0 &    3.4 +3.4/-2.2 & 1.0 & 2.9e-02 \\ 
 17 & 131133.9-343331 & 197.89154 & -34.55871 & 1.07 & 5.6 & 443 & .. &    2.4 +2.9/-1.6 & 0.8 & 2.4e-02 &    2.8 +2.9/-1.6 & 0.9 & 1.4e-03 &   -0.4 +1.9/0.0 & -0.2 & 1.0 \\ 
 18 & 131134.3-342303 & 197.89296 & -34.38430 & 1.14 & 7.1 & 424 & -0.06 &    5.0 +3.8/-2.6 & 1.3 & 5.1e-03 &    2.4 +2.9/-1.6 & 0.8 & 3.1e-02 &    2.6 +3.2/-1.9 & 0.8 & 5.2e-02 \\ 
 19 & 131134.3-342756 & 197.89323 & -34.46568 & 0.18 & 3.7 & 387 & -0.34 &   21.9 +5.8/-4.7 & 3.8 & 0.0 &   14.9 +5.0/-3.8 & 3.0 & 2.5e-30 &    6.9 +3.8/-2.6 & 1.8 & 1.5e-11 \\ 
 20 & 131134.3-342921 & 197.89332 & -34.48929 & 0.45 & 3.5 & 487 & .. &    2.9 +2.9/-1.6 & 1.0 & 2.5e-04 &    3.0 +2.9/-1.6 & 1.0 & 1.5e-05 &   -0.1 +1.9/0.0 & 0.0 & 1.0 \\ 
 21 & 131134.7-343624 & 197.89495 & -34.60690 & 1.00 & 8.0 & 412 & -0.69 &    9.8 +4.6/-3.4 & 2.1 & 4.7e-06 &    1.1 +2.7/-1.3 & 0.4 & 2.3e-01 &    8.7 +4.3/-3.1 & 2.0 & 1.9e-06 \\ 
 22 & 131135.2-342824 & 197.89707 & -34.47335 & 0.37 & 3.4 & 487 & -0.98 &    3.9 +3.2/-1.9 & 1.2 & 1.1e-05 &    4.0 +3.2/-1.9 & 1.2 & 1.9e-07 &   -0.1 +1.9/0.0 & 0.0 & 1.0 \\ 
 23 & 131135.6-342757 & 197.89835 & -34.46605 & 0.39 & 3.5 & 436 & -0.21 &    3.9 +3.2/-1.9 & 1.2 & 1.4e-05 &    3.0 +2.9/-1.6 & 1.0 & 1.9e-05 &    0.9 +2.3/-0.8 & 0.4 & 8.6e-02 \\ 
 24 & 131135.6-342710 & 197.89852 & -34.45300 & 0.52 & 3.8 & 471 & .. &    2.8 +2.9/-1.6 & 1.0 & 6.5e-04 &   -0.1 +1.9/0.0 & 0.0 & 1.0 &    2.9 +2.9/-1.6 & 1.0 & 1.7e-04 \\ 
 25 & 131138.2-342605 & 197.90937 & -34.43479 & 0.50 & 4.1 & 432 & -0.15 &    3.8 +3.2/-1.9 & 1.2 & 7.6e-05 &    0.9 +2.3/-0.8 & 0.4 & 6.4e-02 &    2.9 +2.9/-1.6 & 1.0 & 5.1e-04 \\ 
 26 & 131139.3-342829 & 197.91403 & -34.47494 & 0.23 & 2.6 & 489 & -0.16 &    5.9 +3.6/-2.4 & 1.6 & 2.5e-10 &    4.0 +3.2/-1.9 & 1.2 & 1.2e-08 &    1.9 +2.7/-1.3 & 0.7 & 1.4e-03 \\ 
 27 & 131140.0-343023 & 197.91700 & -34.50657 & 0.30 & 2.6 & 495 & -1.85 &    3.9 +3.2/-1.9 & 1.2 & 1.3e-06 &    2.0 +2.7/-1.3 & 0.8 & 1.7e-04 &    1.9 +2.7/-1.3 & 0.7 & 1.6e-03 \\ 
 28 & 131141.1-343301 & 197.92156 & -34.55055 & 0.66 & 4.4 & 491 & .. &    2.7 +2.9/-1.6 & 0.9 & 3.2e-03 &   -0.1 +1.9/0.0 & -0.1 & 1.0 &    2.8 +2.9/-1.6 & 1.0 & 9.6e-04 \\ 
 29 & 131141.9-342955 & 197.92481 & -34.49881 & 0.19 & 2.1 & 501 & -0.25 &    6.9 +3.8/-2.6 & 1.8 & 4.4e-13 &    1.0 +2.3/-0.8 & 0.4 & 1.8e-02 &    6.0 +3.6/-2.4 & 1.7 & 6.3e-12 \\ 
 30 & 131143.3-342826 & 197.93042 & -34.47410 & 0.27 & 1.8 & 473 & .. &    2.9 +2.9/-1.6 & 1.0 & 2.9e-05 &   0.0 +1.9/0.0 & 0.0 & 1.0 &    3.0 +2.9/-1.6 & 1.0 & 1.0e-05 \\ 
 31 & 131143.7-342749 & 197.93212 & -34.46388 & 0.26 & 2.1 & 497 & -1.48 &    3.9 +3.2/-1.9 & 1.2 & 2.1e-06 &    3.0 +2.9/-1.6 & 1.0 & 7.3e-06 &    0.9 +2.3/-0.8 & 0.4 & 4.9e-02 \\ 
 32 & 131143.9-342500 & 197.93320 & -34.41678 & 0.26 & 4.5 & 425 & -0.40 &   18.7 +5.4/-4.3 & 3.4 & 7.5e-27 &    6.9 +3.8/-2.6 & 1.8 & 7.8e-11 &   11.8 +4.6/-3.4 & 2.6 & 2.0e-17 \\ 
 33 & 131144.6-343100 & 197.93619 & -34.51691 & 0.24 & 2.3 & 500 & -0.28 &    4.9 +3.4/-2.2 & 1.4 & 1.9e-08 &    2.0 +2.7/-1.3 & 0.7 & 3.1e-04 &    3.0 +2.9/-1.6 & 1.0 & 2.1e-05 \\ 
 34 & 131145.6-343313 & 197.94042 & -34.55386 & 0.43 & 4.2 & 447 & -1.66 &    5.7 +3.6/-2.4 & 1.6 & 4.0e-07 &    1.9 +2.7/-1.3 & 0.7 & 3.5e-03 &    3.8 +3.2/-1.9 & 1.2 & 3.9e-05 \\ 
 35 & 131145.7-343030 & 197.94049 & -34.50848 & 0.07 & 1.8 & 215 & -0.22 &   54.0 +8.4/-7.3 & 6.4 & 0.0 &   32.0 +6.7/-5.6 & 4.8 & 0.0 &   22.0 +5.8/-4.7 & 3.8 & 0.0 \\ 
 36 & 131146.0-342424 & 197.94207 & -34.40688 & 0.39 & 4.9 & 433 & -0.03 &   11.6 +4.6/-3.4 & 2.5 & 1.2e-13 &    9.8 +4.3/-3.1 & 2.3 & 2.2e-14 &    1.7 +2.7/-1.3 & 0.7 & 2.9e-02 \\ 
 37 & 131146.3-343259 & 197.94310 & -34.54985 & 0.55 & 3.9 & 452 & .. &    2.8 +2.9/-1.6 & 0.9 & 1.8e-03 &    1.9 +2.7/-1.3 & 0.7 & 3.0e-03 &    0.8 +2.3/-0.8 & 0.4 & 1.5e-01 \\ 
 38 & 131146.4-343050 & 197.94346 & -34.51394 & 0.18 & 1.9 & 506 & -0.02 &    7.9 +4.0/-2.8 & 2.0 & 1.3e-14 &    2.0 +2.7/-1.3 & 0.7 & 2.8e-04 &    6.0 +3.6/-2.4 & 1.7 & 1.3e-11 \\ 
 39 & 131147.0-343205 & 197.94600 & -34.53480 & 0.06 & 3.0 & 430 & -0.37 &  113.9 +11.7/-10.7 & 9.7 & 0.0 &   61.0 +8.9/-7.8 & 6.9 & 0.0 &   52.9 +8.3/-7.3 & 6.3 & 0.0 \\ 
 40 & 131147.0-342953 & 197.94607 & -34.49825 & 0.21 & 1.1 & 504 & -0.21 &    4.0 +3.2/-1.9 & 1.2 & 1.9e-07 &    3.0 +2.9/-1.6 & 1.0 & 4.9e-07 &    1.0 +2.3/-0.8 & 0.4 & 3.2e-02 \\ 
 41 & 131148.6-343729 & 197.95259 & -34.62488 & 1.11 & 8.3 & 445 & -0.04 &    7.7 +4.4/-3.3 & 1.7 & 8.6e-04 &    2.7 +3.2/-1.9 & 0.8 & 4.9e-02 &    5.0 +3.8/-2.6 & 1.3 & 5.6e-03 \\ 
 42 & 131149.0-342829 & 197.95455 & -34.47476 & 0.23 & 0.9 & 406 & .. &    3.0 +2.9/-1.6 & 1.0 & 7.8e-06 &    1.0 +2.3/-0.8 & 0.4 & 1.5e-02 &    2.0 +2.7/-1.3 & 0.7 & 2.3e-04 \\ 
 43 & 131149.2-343817 & 197.95519 & -34.63808 & 0.77 & 9.1 & 432 & -0.32 &   27.3 +6.7/-5.6 & 4.0 & 1.9e-15 &   20.4 +5.8/-4.7 & 3.5 & 3.0e-16 &    6.9 +4.3/-3.1 & 1.6 & 2.0e-03 \\ 
 44 & 131149.4-342607 & 197.95624 & -34.43547 & 0.35 & 3.1 & 445 & -0.94 &    3.9 +3.2/-1.9 & 1.2 & 7.5e-06 &    2.9 +2.9/-1.6 & 1.0 & 2.3e-05 &    0.9 +2.3/-0.8 & 0.4 & 6.5e-02 \\ 
 45 & 131149.6-343330 & 197.95677 & -34.55853 & 0.65 & 4.3 & 493 & .. &    2.7 +2.9/-1.6 & 0.9 & 3.2e-03 &   -0.1 +1.9/0.0 & -0.1 & 1.0 &    2.8 +2.9/-1.6 & 1.0 & 9.7e-04 \\ 
 46 & 131150.1-343743 & 197.95910 & -34.62888 & 1.33 & 8.5 & 441 & -0.06 &    6.4 +4.3/-3.1 & 1.5 & 5.3e-03 &    2.7 +3.2/-1.9 & 0.9 & 4.6e-02 &    3.6 +3.6/-2.4 & 1.0 & 3.8e-02 \\ 
 47 & 131150.5-343029 & 197.96059 & -34.50809 & 0.23 & 1.3 & 483 & -0.04 &    3.9 +3.2/-1.9 & 1.2 & 3.5e-07 &    3.0 +2.9/-1.6 & 1.0 & 1.9e-06 &    1.0 +2.3/-0.8 & 0.4 & 3.2e-02 \\ 
 48 & 131152.1-342432 & 197.96709 & -34.40897 & 0.54 & 4.7 & 428 & -1.17 &    4.6 +3.4/-2.2 & 1.4 & 3.9e-05 &    2.8 +2.9/-1.6 & 1.0 & 6.8e-04 &    1.8 +2.7/-1.3 & 0.7 & 1.7e-02 \\ 
 49 & 131153.3-343212 & 197.97224 & -34.53684 & 0.27 & 3.0 & 475 & -1.44 &    6.9 +3.8/-2.6 & 1.8 & 7.3e-11 &    7.0 +3.8/-2.6 & 1.8 & 1.8e-13 &   -0.1 +1.9/0.0 & 0.0 & 1.0 \\ 
 50 & 131153.5-342528 & 197.97325 & -34.42462 & 0.50 & 3.7 & 468 & .. &    2.8 +2.9/-1.6 & 1.0 & 8.2e-04 &    2.0 +2.7/-1.3 & 0.7 & 1.1e-03 &    0.9 +2.3/-0.8 & 0.4 & 1.2e-01 \\ 
\hline\hline
\end{tabular}
\end{center}
All X-ray sources found in the field of the four ACIS-I chips in the 
\Chandra\ observation of \aJ1311.
\label{table-all1311}
\end{sidewaystable}

\addtocounter{table}{-1}
\begin{sidewaystable}[ht]
\footnotesize
\begin{center}
\caption{({\it continued})}
    \tabcolsep 5.2pt
\begin{tabular}{lcccccccccccccccc}
\hline\hline
\multicolumn{1}{c}{N} &
\multicolumn{1}{c}{CXOU J} &
\multicolumn{1}{c}{R.A.} &
\multicolumn{1}{c}{Decl.} &
\multicolumn{1}{c}{$r$} &
\multicolumn{1}{c}{$D$} &
\multicolumn{1}{c}{$A_{\rm eff}$} &
\multicolumn{1}{c}{log} &
\multicolumn{1}{c}{Counts} &
\multicolumn{1}{c}{$\sigma$} &
\multicolumn{1}{c}{$P_{\rm B}$} &
\multicolumn{1}{c}{Counts} &
\multicolumn{1}{c}{$\sigma$} &
\multicolumn{1}{c}{$P_{\rm B}$} &
\multicolumn{1}{c}{Counts} &
\multicolumn{1}{c}{$\sigma$} &
\multicolumn{1}{c}{$P_{\rm B}$} \\
\cline{9-11}\cline{12-14}\cline{15-17}
\multicolumn{1}{c}{} &
\multicolumn{1}{c}{Name} &
\multicolumn{1}{c}{(deg)} &
\multicolumn{1}{c}{(deg)} &
\multicolumn{1}{c}{($\arcsec$)} &
\multicolumn{1}{c}{($'$)} &
\multicolumn{1}{c}{(cm$^2$)} &
\multicolumn{1}{c}{$P_{\rm KS}$} &
\multicolumn{3}{c}{($0.5-8$ keV)} &
\multicolumn{3}{c}{($0.5-2$ keV)} &
\multicolumn{3}{c}{($2-8$ keV)} \\
\hline
 51 & 131154.3-343224 & 197.97630 & -34.54024 & 0.46 & 3.3 & 434 & .. &    2.9 +2.9/-1.6 & 1.0 & 3.9e-04 &    0.9 +2.3/-0.8 & 0.4 & 5.5e-02 &    1.9 +2.7/-1.3 & 0.7 & 3.2e-03 \\ 
 52 & 131154.3-342434 & 197.97658 & -34.40946 & 0.68 & 4.7 & 237 & .. &    2.8 +2.9/-1.6 & 1.0 & 1.0e-03 &    0.9 +2.3/-0.8 & 0.4 & 7.8e-02 &    1.9 +2.7/-1.3 & 0.7 & 5.8e-03 \\ 
 53 & 131154.5-342719 & 197.97720 & -34.45538 & 0.29 & 2.0 & 477 & .. &    2.9 +2.9/-1.6 & 1.0 & 6.4e-05 &   0.0 +1.9/0.0 & 0.0 & 1.0 &    3.0 +2.9/-1.6 & 1.0 & 1.6e-05 \\ 
 54 & 131154.9-342642 & 197.97910 & -34.44522 & 0.34 & 2.6 & 483 & .. &    2.9 +2.9/-1.6 & 1.0 & 7.8e-05 &    2.0 +2.7/-1.3 & 0.7 & 4.9e-04 &    1.0 +2.3/-0.8 & 0.4 & 4.7e-02 \\ 
 55 & 131156.1-342126 & 197.98379 & -34.35747 & 1.22 & 7.8 & 386 & -0.71 &    6.2 +4.0/-2.8 & 1.6 & 6.2e-04 &    4.3 +3.4/-2.2 & 1.3 & 1.2e-03 &    2.0 +2.9/-1.6 & 0.7 & 9.3e-02 \\ 
 56 & 131156.2-342335 & 197.98440 & -34.39309 & 0.30 & 5.7 & 427 & -0.32 &   29.2 +6.5/-5.4 & 4.5 & 9.8e-35 &   19.8 +5.6/-4.4 & 3.6 & 1.4e-29 &    9.4 +4.3/-3.1 & 2.2 & 1.1e-09 \\ 
 57 & 131156.6-342055 & 197.98587 & -34.34866 & 0.49 & 8.3 & 252 & -0.04 &   20.3 +5.8/-4.7 & 3.5 & 3.4e-13 &   17.2 +5.3/-4.2 & 3.2 & 3.6e-13 &    3.1 +3.2/-1.9 & 1.0 & 2.8e-02 \\ 
 58 & 131156.7-343045 & 197.98631 & -34.51257 & 0.12 & 1.9 & 507 & -0.11 &   18.9 +5.4/-4.3 & 3.5 & 0.0 &   13.0 +4.7/-3.6 & 2.8 & 1.6e-30 &    6.0 +3.6/-2.4 & 1.6 & 2.4e-11 \\ 
 59 & 131156.8-342403 & 197.98691 & -34.40101 & 1.17 & 5.3 & 457 & .. &    1.4 +2.7/-1.3 & 0.5 & 1.2e-01 &    1.8 +2.7/-1.3 & 0.7 & 2.3e-02 &   -0.4 +1.9/0.0 & -0.2 & 1.0 \\ 
 60 & 131157.1-342055 & 197.98814 & -34.34882 & 0.91 & 8.4 & 166 & .. &    1.8 +2.9/-1.6 & 0.6 & 1.3e-01 &    0.2 +2.3/-0.8 & 0.1 & 5.5e-01 &    1.6 +2.7/-1.3 & 0.6 & 7.7e-02 \\ 
 61 & 131158.2-342734 & 197.99275 & -34.45946 & 0.19 & 2.2 & 458 & -0.75 &    6.9 +3.8/-2.6 & 1.8 & 1.2e-12 &    4.0 +3.2/-1.9 & 1.2 & 1.6e-08 &    3.0 +2.9/-1.6 & 1.0 & 1.3e-05 \\ 
 62 & 131159.2-342955 & 197.99670 & -34.49864 & 0.25 & 1.8 & 457 & -1.82 &    3.9 +3.2/-1.9 & 1.2 & 5.5e-07 &    4.0 +3.2/-1.9 & 1.2 & 8.4e-09 &   0.0 +1.9/0.0 & 0.0 & 1.0 \\ 
 63 & 131159.2-343133 & 197.99692 & -34.52606 & 0.22 & 2.9 & 465 & -0.02 &    9.9 +4.3/-3.1 & 2.3 & 2.9e-17 &    1.0 +2.3/-0.8 & 0.4 & 4.0e-02 &    8.9 +4.1/-2.9 & 2.2 & 3.2e-17 \\ 
 64 & 131159.5-343555 & 197.99810 & -34.59878 & 0.35 & 6.9 & 424 & -0.02 &   52.4 +8.4/-7.3 & 6.2 & 0.0 &   35.4 +7.1/-6.0 & 5.0 & 0.0 &   17.0 +5.3/-4.2 & 3.2 & 6.4e-16 \\ 
 65 & 131200.4-343231 & 198.00170 & -34.54213 & 0.17 & 3.8 & 461 & -0.00 &   33.7 +6.9/-5.8 & 4.9 & 0.0 &   22.9 +5.9/-4.8 & 3.9 & 0.0 &   10.8 +4.4/-3.3 & 2.4 & 4.9e-16 \\ 
 66 & 131202.0-343234 & 198.00857 & -34.54281 & 0.22 & 4.0 & 459 & -0.03 &   22.7 +5.9/-4.8 & 3.9 & 7.7e-35 &    5.9 +3.6/-2.4 & 1.6 & 1.7e-09 &   16.8 +5.2/-4.1 & 3.2 & 7.0e-27 \\ 
 67 & 131202.5-342835 & 198.01065 & -34.47657 & 0.34 & 2.4 & 461 & .. &    2.9 +2.9/-1.6 & 1.0 & 1.3e-04 &    2.0 +2.7/-1.3 & 0.7 & 8.1e-04 &    0.9 +2.3/-0.8 & 0.4 & 5.2e-02 \\ 
 68 & 131202.9-342539 & 198.01213 & -34.42761 & 0.52 & 4.3 & 473 & -0.01 &    3.8 +3.2/-1.9 & 1.2 & 1.0e-04 &    2.9 +2.9/-1.6 & 1.0 & 7.6e-05 &    0.8 +2.3/-0.8 & 0.4 & 1.4e-01 \\ 
 69 & 131203.9-342026 & 198.01630 & -34.34067 & 0.76 & 9.1 & 372 & -0.76 &   27.1 +6.6/-5.5 & 4.1 & 1.4e-16 &   27.9 +6.5/-5.4 & 4.3 & 4.5e-27 &   -0.9 +2.7/-1.3 & -0.3 & 7.8e-01 \\ 
 70 & 131203.9-343653 & 198.01636 & -34.61482 & 0.59 & 8.1 & 437 & -0.11 &   33.4 +7.2/-6.1 & 4.7 & 7.2e-23 &   18.9 +5.6/-4.4 & 3.4 & 6.9e-17 &   14.5 +5.2/-4.1 & 2.8 & 6.2e-09 \\ 
 71 & 131204.4-343559 & 198.01873 & -34.59985 & 1.11 & 7.3 & 451 & -0.27 &    5.9 +4.0/-2.8 & 1.5 & 1.9e-03 &   -0.7 +1.9/0.0 & -0.4 & 1.0 &    6.6 +4.0/-2.8 & 1.7 & 1.5e-04 \\ 
 72 & 131205.2-342502 & 198.02176 & -34.41736 & 0.67 & 5.0 & 456 & -0.01 &    3.6 +3.2/-1.9 & 1.1 & 5.2e-04 &    0.9 +2.3/-0.8 & 0.4 & 1.3e-01 &    2.8 +2.9/-1.6 & 0.9 & 1.6e-03 \\ 
 73 & 131205.3-342813 & 198.02211 & -34.47039 & 0.36 & 3.0 & 485 & -0.68 &    3.9 +3.2/-1.9 & 1.2 & 1.2e-05 &    2.9 +2.9/-1.6 & 1.0 & 3.3e-05 &    0.9 +2.3/-0.8 & 0.4 & 7.3e-02 \\ 
 74 & 131205.6-343126 & 198.02371 & -34.52414 & 0.53 & 3.7 & 490 & .. &    2.8 +2.9/-1.6 & 0.9 & 1.8e-03 &   -0.1 +1.9/0.0 & 0.0 & 1.0 &    2.8 +2.9/-1.6 & 1.0 & 5.7e-04 \\ 
 75 & 131206.1-343152 & 198.02560 & -34.53122 & 0.20 & 4.0 & 482 & -0.27 &   23.8 +6.0/-4.9 & 4.0 & 5.8e-38 &   16.9 +5.2/-4.1 & 3.2 & 1.8e-31 &    6.8 +3.8/-2.6 & 1.8 & 6.2e-10 \\ 
 76 & 131206.2-343449 & 198.02606 & -34.58041 & 0.44 & 6.4 & 457 & -1.40 &   26.7 +6.4/-5.3 & 4.2 & 3.1e-26 &   17.6 +5.3/-4.2 & 3.3 & 2.0e-21 &    9.1 +4.3/-3.1 & 2.1 & 4.8e-08 \\ 
 77 & 131206.5-342910 & 198.02731 & -34.48618 & 0.43 & 3.1 & 433 & .. &    2.9 +2.9/-1.6 & 1.0 & 2.3e-04 &    2.0 +2.7/-1.3 & 0.7 & 9.0e-04 &    0.9 +2.3/-0.8 & 0.4 & 6.9e-02 \\ 
 78 & 131207.4-343107 & 198.03086 & -34.51887 & 0.20 & 3.8 & 486 & -0.21 &   20.8 +5.7/-4.5 & 3.7 & 2.1e-34 &   13.9 +4.8/-3.7 & 2.9 & 3.6e-27 &    6.9 +3.8/-2.6 & 1.8 & 1.9e-10 \\ 
 79 & 131207.7-342622 & 198.03237 & -34.43961 & 0.65 & 4.4 & 445 & .. &    2.8 +2.9/-1.6 & 0.9 & 1.4e-03 &    1.9 +2.7/-1.3 & 0.7 & 2.7e-03 &    0.9 +2.3/-0.8 & 0.4 & 1.3e-01 \\ 
 80 & 131207.8-342938 & 198.03253 & -34.49399 & 0.47 & 3.4 & 483 & .. &    2.9 +2.9/-1.6 & 1.0 & 3.5e-04 &    1.0 +2.3/-0.8 & 0.4 & 4.8e-02 &    1.9 +2.7/-1.3 & 0.7 & 3.3e-03 \\ 
 81 & 131209.2-343622 & 198.03871 & -34.60625 & 0.58 & 8.1 & 440 & -0.10 &   34.2 +7.1/-6.1 & 4.8 & 9.1e-27 &   23.3 +6.0/-4.9 & 3.9 & 4.1e-25 &   11.0 +4.7/-3.6 & 2.3 & 5.2e-07 \\ 
 82 & 131209.8-343440 & 198.04118 & -34.57790 & 0.80 & 6.7 & 437 & -0.12 &    8.8 +4.3/-3.1 & 2.0 & 9.0e-07 &    7.5 +4.0/-2.8 & 1.9 & 6.5e-08 &    1.2 +2.7/-1.3 & 0.5 & 1.8e-01 \\ 
 83 & 131211.7-342558 & 198.04886 & -34.43281 & 0.68 & 5.3 & 424 & -0.35 &    4.6 +3.4/-2.2 & 1.4 & 5.8e-05 &   -0.2 +1.9/0.0 & -0.1 & 1.0 &    4.8 +3.4/-2.2 & 1.4 & 4.3e-06 \\ 
 84 & 131217.8-342938 & 198.07458 & -34.49393 & 0.76 & 5.5 & 470 & -0.12 &    4.3 +3.4/-2.2 & 1.3 & 6.2e-04 &    3.8 +3.2/-1.9 & 1.2 & 1.1e-04 &    0.6 +2.3/-0.8 & 0.2 & 3.5e-01 \\ 
 85 & 131218.4-342819 & 198.07683 & -34.47197 & 0.57 & 5.6 & 373 & -0.05 &    9.4 +4.3/-3.1 & 2.2 & 1.2e-09 &    6.7 +3.8/-2.6 & 1.8 & 2.0e-08 &    2.7 +2.9/-1.6 & 0.9 & 5.1e-03 \\ 
 86 & 131219.5-342811 & 198.08141 & -34.46998 & 1.01 & 5.9 & 349 & -0.45 &    3.4 +3.2/-1.9 & 1.1 & 4.1e-03 &    0.8 +2.3/-0.8 & 0.3 & 2.2e-01 &    2.6 +2.9/-1.6 & 0.9 & 7.0e-03 \\ 
 87 & 131220.2-342931 & 198.08456 & -34.49205 & 1.01 & 6.0 & 459 & -0.10 &    3.2 +3.2/-1.9 & 1.0 & 1.0e-02 &    1.7 +2.7/-1.3 & 0.6 & 4.5e-02 &    1.5 +2.7/-1.3 & 0.6 & 8.8e-02 \\ 
 88 & 131220.4-343456 & 198.08521 & -34.58249 & 1.53 & 8.3 & 404 & -1.60 &    4.1 +3.8/-2.6 & 1.1 & 3.0e-02 &    4.1 +3.4/-2.2 & 1.2 & 3.4e-03 &    0.1 +2.7/-1.3 & 0.0 & 5.8e-01 \\ 
 89 & 131223.1-342912 & 198.09643 & -34.48689 & 0.34 & 6.5 & 448 & -1.70 &   39.8 +7.5/-6.4 & 5.3 & 0.0 &   29.5 +6.5/-5.4 & 4.5 & 0.0 &   10.3 +4.4/-3.3 & 2.3 & 5.4e-10 \\ 
 90 & 131224.5-342859 & 198.10229 & -34.48330 & 0.93 & 6.8 & 449 & -0.04 &    6.5 +4.0/-2.8 & 1.6 & 2.0e-04 &    3.4 +3.2/-1.9 & 1.1 & 4.2e-03 &    3.1 +3.2/-1.9 & 1.0 & 1.4e-02 \\ 
 91 & 131225.3-342501 & 198.10569 & -34.41703 & 0.78 & 8.1 & 404 & -0.81 &   19.7 +5.8/-4.7 & 3.4 & 6.4e-14 &   19.0 +5.6/-4.4 & 3.4 & 3.2e-18 &    0.6 +2.7/-1.3 & 0.2 & 4.0e-01 \\ 
 92 & 131225.5-342628 & 198.10644 & -34.44130 & 0.66 & 7.5 & 410 & -0.84 &   20.2 +5.8/-4.7 & 3.5 & 4.4e-16 &   13.2 +4.8/-3.7 & 2.7 & 6.2e-13 &    6.9 +4.0/-2.8 & 1.8 & 2.3e-05 \\ 
 93 & 131226.8-343029 & 198.11173 & -34.50825 & 1.26 & 7.4 & 411 & -0.40 &    4.2 +3.6/-2.4 & 1.2 & 1.1e-02 &    2.3 +2.9/-1.6 & 0.8 & 3.5e-02 &    1.9 +2.9/-1.6 & 0.7 & 1.0e-01 \\ 
 94 & 131228.3-343136 & 198.11798 & -34.52668 & 1.62 & 8.0 & 400 & -1.41 &    2.9 +3.4/-2.2 & 0.9 & 6.3e-02 &    4.2 +3.4/-2.2 & 1.2 & 2.1e-03 &   -1.2 +1.9/0.0 & -0.7 & 1.0 \\ 
 95 & 131229.1-342641 & 198.12128 & -34.44483 & 0.56 & 8.2 & 236 & -1.72 &   37.5 +7.3/-6.2 & 5.1 & 3.4e-38 &   20.3 +5.7/-4.5 & 3.6 & 2.6e-22 &   17.2 +5.3/-4.2 & 3.2 & 1.5e-17 \\ 
 96 & 131234.9-342925 & 198.14571 & -34.49055 & 0.80 & 9.0 & 396 & -0.48 &   24.5 +6.4/-5.3 & 3.8 & 3.5e-15 &   20.8 +5.8/-4.7 & 3.6 & 2.8e-18 &    3.7 +3.6/-2.4 & 1.0 & 3.6e-02 \\ 
 97 & 131236.3-343004 & 198.15137 & -34.50131 & 0.83 & 9.3 & 416 & -0.19 &   25.4 +6.5/-5.4 & 3.9 & 7.9e-16 &    5.0 +3.6/-2.4 & 1.4 & 9.5e-04 &   20.4 +5.9/-4.8 & 3.5 & 1.4e-13 \\ 
\hline\hline
\end{tabular}
\end{center}
\end{sidewaystable}

\begin{sidewaystable}[ht]
\footnotesize
\begin{center}
\caption{}
    \tabcolsep 4.2pt
\begin{tabular}{lcccccccccccccccc}
\hline\hline
\multicolumn{1}{c}{N} &
\multicolumn{1}{c}{CXOU J} &
\multicolumn{1}{c}{R.A.} &
\multicolumn{1}{c}{Decl.} &
\multicolumn{1}{c}{$r$} &
\multicolumn{1}{c}{$D$} &
\multicolumn{1}{c}{$A_{\rm eff}$} &
\multicolumn{1}{c}{log} &
\multicolumn{1}{c}{Counts} &
\multicolumn{1}{c}{$\sigma$} &
\multicolumn{1}{c}{$P_{\rm B}$} &
\multicolumn{1}{c}{Counts} &
\multicolumn{1}{c}{$\sigma$} &
\multicolumn{1}{c}{$P_{\rm B}$} &
\multicolumn{1}{c}{Counts} &
\multicolumn{1}{c}{$\sigma$} &
\multicolumn{1}{c}{$P_{\rm B}$} \\
\cline{9-11}\cline{12-14}\cline{15-17}
\multicolumn{1}{c}{} &
\multicolumn{1}{c}{Name} &
\multicolumn{1}{c}{(deg)} &
\multicolumn{1}{c}{(deg)} &
\multicolumn{1}{c}{($\arcsec$)} &
\multicolumn{1}{c}{($'$)} &
\multicolumn{1}{c}{(cm$^2$)} &
\multicolumn{1}{c}{$P_{\rm KS}$} &
\multicolumn{3}{c}{($0.5-8$ keV)} &
\multicolumn{3}{c}{($0.5-2$ keV)} &
\multicolumn{3}{c}{($2-8$ keV)} \\
\hline
  1 & 165305.4-020127 & 253.27286 &  -2.02433 & 1.84 & 10.2 & 394 & -0.13 &    5.1 +4.5/-3.3 & 1.1 & 4.3e-02 &    0.7 +3.0/-1.7 & 0.2 & 4.2e-01 &    4.4 +4.0/-2.8 & 1.1 & 3.3e-02 \\ 
  2 & 165307.6-015954 & 253.28203 &  -1.99854 & 0.86 & 9.2 & 286 & -0.06 &   22.3 +6.2/-5.1 & 3.6 & 4.3e-13 &    7.3 +4.1/-2.9 & 1.8 & 1.1e-04 &   14.9 +5.2/-4.1 & 2.9 & 5.2e-10 \\ 
  3 & 165309.0-020138 & 253.28751 &  -2.02737 & 0.68 & 9.4 & 410 & -0.10 &   40.3 +7.9/-6.8 & 5.1 & 2.7e-24 &   26.7 +6.5/-5.4 & 4.1 & 2.9e-20 &   13.6 +5.2/-4.1 & 2.6 & 3.1e-07 \\ 
  4 & 165309.7-015915 & 253.29050 &  -1.98769 & 0.80 & 8.6 & 411 & -0.21 &   20.4 +6.0/-4.9 & 3.4 & 4.6e-12 &   14.6 +5.1/-4.0 & 2.9 & 1.5e-11 &    5.8 +4.0/-2.8 & 1.5 & 2.6e-03 \\ 
  5 & 165310.3-015926 & 253.29301 &  -1.99058 & 0.99 & 8.5 & 373 & -0.29 &    9.7 +4.6/-3.4 & 2.1 & 7.5e-06 &    2.1 +2.9/-1.6 & 0.7 & 6.4e-02 &    7.6 +4.1/-2.9 & 1.8 & 2.4e-05 \\ 
  6 & 165312.4-015943 & 253.30206 &  -1.99541 & 1.25 & 8.0 & 422 & -0.21 &    6.1 +4.1/-3.0 & 1.5 & 4.1e-03 &    3.9 +3.4/-2.2 & 1.2 & 5.9e-03 &    2.1 +3.2/-1.9 & 0.7 & 1.3e-01 \\ 
  7 & 165315.2-020314 & 253.31349 &  -2.05407 & 0.96 & 8.9 & 420 & -0.18 &   14.2 +5.3/-4.2 & 2.7 & 2.9e-07 &    5.6 +3.8/-2.6 & 1.5 & 1.1e-03 &    8.6 +4.4/-3.3 & 1.9 & 6.7e-05 \\ 
  8 & 165315.6-015822 & 253.31502 &  -1.97283 & 0.06 & 7.1 & 360 & -0.12 & 1674.0 +42.0/-40.9 & 39.9 & 0.0 & 1138.4 +34.8/-33.8 & 32.7 & 0.0 &  535.6 +24.2/-23.2 & 22.1 & 0.0 \\ 
  9 & 165318.2-020116 & 253.32620 &  -2.02115 & 1.15 & 7.2 & 436 & -0.16 &    4.3 +3.6/-2.4 & 1.2 & 8.4e-03 &    1.4 +2.7/-1.3 & 0.5 & 1.1e-01 &    2.9 +3.2/-1.9 & 0.9 & 3.0e-02 \\ 
 10 & 165318.3-015927 & 253.32629 &  -1.99097 & 1.08 & 6.5 & 444 & -1.41 &    3.6 +3.4/-2.2 & 1.1 & 1.5e-02 &   -0.5 +1.9/0.0 & -0.3 & 1.0 &    4.1 +3.4/-2.2 & 1.2 & 2.6e-03 \\ 
 11 & 165318.4-015434 & 253.32693 &  -1.90966 & 0.92 & 7.2 & 394 & -0.64 &    9.6 +4.6/-3.4 & 2.1 & 1.2e-05 &    6.0 +3.8/-2.6 & 1.6 & 1.3e-04 &    3.6 +3.4/-2.2 & 1.1 & 1.6e-02 \\ 
 12 & 165321.5-015703 & 253.33975 &  -1.95102 & 0.75 & 5.7 & 393 & -0.07 &    5.0 +3.6/-2.4 & 1.4 & 6.1e-04 &   -0.5 +1.9/0.0 & -0.2 & 1.0 &    5.5 +3.6/-2.4 & 1.5 & 2.2e-05 \\ 
 13 & 165323.1-015252 & 253.34642 &  -1.88130 & 1.40 & 7.3 & 379 & -0.07 &    3.1 +3.4/-2.2 & 0.9 & 4.3e-02 &    3.4 +3.2/-1.9 & 1.1 & 4.5e-03 &   -0.2 +2.3/-0.8 & -0.1 & 7.1e-01 \\ 
 14 & 165323.2-020451 & 253.34676 &  -2.08089 & 0.38 & 8.6 & 335 & -1.42 &   89.4 +10.6/-9.6 & 8.4 & 0.0 &   59.2 +8.8/-7.7 & 6.7 & 0.0 &   30.2 +6.7/-5.6 & 4.5 & 8.8e-27 \\ 
 15 & 165323.2-015809 & 253.34687 &  -1.96928 & 0.60 & 5.1 & 459 & -0.26 &    5.4 +3.6/-2.4 & 1.5 & 3.3e-05 &    1.8 +2.7/-1.3 & 0.7 & 2.1e-02 &    3.6 +3.2/-1.9 & 1.1 & 5.3e-04 \\ 
 16 & 165323.9-020155 & 253.34988 &  -2.03222 & 1.13 & 6.3 & 454 & .. &    2.0 +2.9/-1.6 & 0.7 & 8.1e-02 &    2.7 +2.9/-1.6 & 0.9 & 5.0e-03 &   -0.7 +1.9/0.0 & -0.4 & 1.0 \\ 
 17 & 165324.6-015932 & 253.35291 &  -1.99234 & 0.40 & 5.0 & 375 & -0.03 &   10.5 +4.4/-3.3 & 2.4 & 4.8e-12 &    6.8 +3.8/-2.6 & 1.8 & 1.1e-09 &    3.7 +3.2/-1.9 & 1.2 & 2.6e-04 \\ 
 18 & 165325.7-015617 & 253.35745 &  -1.93829 & 0.71 & 4.8 & 343 & -0.43 &    3.5 +3.2/-1.9 & 1.1 & 1.4e-03 &   -0.2 +1.9/0.0 & -0.1 & 1.0 &    3.7 +3.2/-1.9 & 1.2 & 2.9e-04 \\ 
 19 & 165325.8-015107 & 253.35787 &  -1.85215 & 0.70 & 8.2 & 357 & -1.04 &   23.4 +6.2/-5.1 & 3.8 & 1.3e-16 &    8.9 +4.3/-3.1 & 2.1 & 5.9e-07 &   14.5 +5.1/-4.0 & 2.8 & 3.9e-11 \\ 
 20 & 165325.9-015316 & 253.35808 &  -1.88782 & 0.83 & 6.5 & 366 & -0.06 &    6.8 +4.0/-2.8 & 1.7 & 5.8e-05 &    4.7 +3.4/-2.2 & 1.4 & 3.6e-05 &    2.1 +2.9/-1.6 & 0.7 & 6.8e-02 \\ 
 21 & 165326.3-015502 & 253.35973 &  -1.91743 & 0.84 & 5.3 & 417 & -0.14 &    3.3 +3.2/-1.9 & 1.0 & 7.2e-03 &    1.8 +2.7/-1.3 & 0.7 & 1.9e-02 &    1.5 +2.7/-1.3 & 0.6 & 1.0e-01 \\ 
 22 & 165326.9-015741 & 253.36237 &  -1.96154 & 0.61 & 4.2 & 465 & .. &    2.7 +2.9/-1.6 & 0.9 & 4.0e-03 &    1.9 +2.7/-1.3 & 0.7 & 5.3e-03 &    0.8 +2.3/-0.8 & 0.3 & 1.9e-01 \\ 
 23 & 165327.4-015118 & 253.36437 &  -1.85509 & 1.67 & 7.9 & 423 & -0.07 &    3.6 +3.6/-2.4 & 1.0 & 3.8e-02 &    4.0 +3.4/-2.2 & 1.2 & 4.8e-03 &   -0.4 +2.3/-0.8 & -0.2 & 7.5e-01 \\ 
 24 & 165328.4-020009 & 253.36843 &  -2.00261 & 0.28 & 4.4 & 471 & -0.02 &   13.7 +4.8/-3.7 & 2.8 & 1.7e-18 &   11.9 +4.6/-3.4 & 2.6 & 2.5e-19 &    1.8 +2.7/-1.3 & 0.7 & 1.5e-02 \\ 
 25 & 165330.0-015827 & 253.37516 &  -1.97442 & 0.45 & 3.5 & 479 & .. &    2.8 +2.9/-1.6 & 1.0 & 6.6e-04 &    1.9 +2.7/-1.3 & 0.7 & 3.1e-03 &    0.9 +2.3/-0.8 & 0.4 & 8.1e-02 \\ 
 26 & 165331.4-020137 & 253.38113 &  -2.02713 & 0.73 & 4.8 & 462 & .. &    2.6 +2.9/-1.6 & 0.9 & 6.8e-03 &   -0.2 +1.9/0.0 & -0.1 & 1.0 &    2.8 +2.9/-1.6 & 0.9 & 1.3e-03 \\ 
 27 & 165332.8-015902 & 253.38701 &  -1.98411 & 0.37 & 2.9 & 486 & .. &    2.9 +2.9/-1.6 & 1.0 & 3.1e-04 &    0.9 +2.3/-0.8 & 0.4 & 5.1e-02 &    1.9 +2.7/-1.3 & 0.7 & 2.7e-03 \\ 
 28 & 165333.5-015259 & 253.38978 &  -1.88320 & 0.12 & 5.6 & 451 & -0.06 &  173.3 +14.2/-13.2 & 12.2 & 0.0 &  172.8 +14.2/-13.1 & 12.2 & 0.0 &    0.6 +2.3/-0.8 & 0.2 & 3.6e-01 \\ 
 29 & 165334.5-020230 & 253.39413 &  -2.04180 & 0.38 & 5.1 & 476 & -0.34 &   13.5 +4.8/-3.7 & 2.8 & 2.5e-15 &    7.8 +4.0/-2.8 & 2.0 & 1.7e-10 &    5.7 +3.6/-2.4 & 1.6 & 1.6e-06 \\ 
 30 & 165334.8-020031 & 253.39522 &  -2.00879 & 0.43 & 3.4 & 491 & .. &    2.9 +2.9/-1.6 & 1.0 & 4.8e-04 &   -0.1 +1.9/0.0 & 0.0 & 1.0 &    2.9 +2.9/-1.6 & 1.0 & 1.1e-04 \\ 
 31 & 165334.8-015803 & 253.39522 &  -1.96760 & 0.23 & 2.2 & 485 & -0.31 &    4.9 +3.4/-2.2 & 1.4 & 1.2e-08 &    3.0 +2.9/-1.6 & 1.0 & 2.1e-06 &    2.0 +2.7/-1.3 & 0.7 & 1.0e-03 \\ 
 32 & 165334.9-020317 & 253.39542 &  -2.05488 & 0.87 & 5.7 & 470 & -0.25 &    4.1 +3.4/-2.2 & 1.2 & 2.7e-03 &    0.6 +2.3/-0.8 & 0.3 & 3.3e-01 &    3.5 +3.2/-1.9 & 1.1 & 2.2e-03 \\ 
 33 & 165336.3-014918 & 253.40146 &  -1.82192 & 0.36 & 8.9 & 358 & -0.06 &  101.7 +11.3/-10.3 & 9.0 & 0.0 &   68.9 +9.5/-8.4 & 7.3 & 0.0 &   32.8 +7.0/-5.9 & 4.7 & 3.4e-26 \\ 
 34 & 165336.7-014907 & 253.40294 &  -1.81872 & 0.60 & 9.1 & 216 & -0.01 &   14.5 +5.1/-4.0 & 2.9 & 9.5e-12 &    7.3 +4.0/-2.8 & 1.8 & 1.3e-06 &    7.3 +4.0/-2.8 & 1.8 & 1.6e-06 \\ 
 35 & 165337.0-020211 & 253.40424 &  -2.03646 & 0.49 & 4.5 & 481 & -0.77 &    5.6 +3.6/-2.4 & 1.6 & 3.3e-06 &    4.8 +3.4/-2.2 & 1.4 & 8.3e-07 &    0.8 +2.3/-0.8 & 0.3 & 2.0e-01 \\ 
 36 & 165337.1-020442 & 253.40491 &  -2.07841 & 0.76 & 6.9 & 420 & -0.16 &   10.5 +4.6/-3.4 & 2.3 & 1.2e-07 &    4.4 +3.4/-2.2 & 1.3 & 5.1e-04 &    6.1 +3.8/-2.6 & 1.6 & 6.1e-05 \\ 
 37 & 165337.2-020020 & 253.40519 &  -2.00568 & 0.18 & 2.9 & 491 & -1.92 &   11.9 +4.6/-3.4 & 2.6 & 3.8e-21 &    5.0 +3.4/-2.2 & 1.5 & 1.4e-09 &    6.9 +3.8/-2.6 & 1.8 & 6.6e-13 \\ 
 38 & 165338.0-015836 & 253.40855 &  -1.97685 & 0.03 & 1.6 & 500 & -0.37 &  306.9 +18.5/-17.5 & 16.6 & 0.0 &  198.0 +15.1/-14.1 & 13.1 & 0.0 &  109.0 +11.5/-10.4 & 9.5 & 0.0 \\ 
 39 & 165338.9-015007 & 253.41248 &  -1.83532 & 1.09 & 8.0 & 330 & -0.43 &    8.0 +4.3/-3.1 & 1.9 & 6.1e-05 &   -0.9 +1.9/0.0 & -0.5 & 1.0 &    8.9 +4.3/-3.1 & 2.1 & 6.0e-07 \\ 
 40 & 165339.1-015412 & 253.41314 &  -1.90339 & 0.54 & 4.0 & 469 & .. &    2.8 +2.9/-1.6 & 0.9 & 1.4e-03 &   -0.1 +1.9/0.0 & 0.0 & 1.0 &    2.9 +2.9/-1.6 & 1.0 & 4.6e-04 \\ 
 41 & 165339.4-015051 & 253.41434 &  -1.84757 & 0.98 & 7.2 & 361 & -0.79 &    6.5 +4.0/-2.8 & 1.6 & 1.8e-04 &    5.4 +3.6/-2.4 & 1.5 & 4.5e-05 &    1.1 +2.7/-1.3 & 0.4 & 2.3e-01 \\ 
 42 & 165339.6-015033 & 253.41537 &  -1.84275 & 0.83 & 7.5 & 339 & -0.73 &   10.5 +4.6/-3.4 & 2.3 & 8.8e-08 &    5.4 +3.6/-2.4 & 1.5 & 4.9e-05 &    5.1 +3.6/-2.4 & 1.4 & 3.7e-04 \\ 
 43 & 165340.3-015329 & 253.41809 &  -1.89151 & 0.69 & 4.6 & 464 & .. &    2.6 +2.9/-1.6 & 0.9 & 6.3e-03 &    2.9 +2.9/-1.6 & 1.0 & 5.6e-04 &   -0.2 +1.9/0.0 & -0.1 & 1.0 \\ 
 44 & 165341.4-015927 & 253.42251 &  -1.99084 & 0.10 & 1.6 & 504 & -1.12 &   18.9 +5.4/-4.3 & 3.5 & 0.0 &   17.0 +5.2/-4.1 & 3.3 & 0.0 &    2.0 +2.7/-1.3 & 0.7 & 8.0e-04 \\ 
 45 & 165341.4-020451 & 253.42281 &  -2.08095 & 1.26 & 6.9 & 460 & -0.10 &    3.6 +3.4/-2.2 & 1.1 & 1.6e-02 &    3.5 +3.2/-1.9 & 1.1 & 2.8e-03 &    0.1 +2.3/-0.8 & 0.1 & 5.8e-01 \\ 
 46 & 165341.6-015425 & 253.42350 &  -1.90709 & 0.42 & 3.6 & 478 & -0.47 &    3.8 +3.2/-1.9 & 1.2 & 5.4e-05 &    0.9 +2.3/-0.8 & 0.4 & 7.3e-02 &    2.9 +2.9/-1.6 & 1.0 & 2.8e-04 \\ 
 47 & 165342.4-020034 & 253.42708 &  -2.00949 & 0.36 & 2.6 & 223 & .. &    3.0 +2.9/-1.6 & 1.0 & 1.2e-05 &    2.0 +2.7/-1.3 & 0.8 & 1.5e-04 &    1.0 +2.3/-0.8 & 0.4 & 2.4e-02 \\ 
 48 & 165342.6-020516 & 253.42769 &  -2.08805 & 1.25 & 7.3 & 418 & -0.18 &    4.3 +3.6/-2.4 & 1.2 & 9.0e-03 &    4.4 +3.4/-2.2 & 1.3 & 5.4e-04 &   -0.1 +2.3/-0.8 & 0.0 & 6.7e-01 \\ 
 49 & 165342.8-020144 & 253.42846 &  -2.02914 & 0.38 & 3.7 & 452 & -0.35 &    5.8 +3.6/-2.4 & 1.6 & 1.1e-07 &    5.9 +3.6/-2.4 & 1.6 & 2.4e-10 &   -0.1 +1.9/0.0 & -0.1 & 1.0 \\ 
 50 & 165343.0-015238 & 253.42925 &  -1.87735 & 0.26 & 5.4 & 463 & -0.33 &   32.4 +6.8/-5.7 & 4.8 & 0.0 &   21.7 +5.8/-4.7 & 3.8 & 2.3e-32 &   10.6 +4.4/-3.3 & 2.4 & 9.4e-13 \\ 
\hline\hline
\end{tabular}
\end{center}
All X-ray sources found in the field of the four ACIS-I chips in the 
\Chandra\ observation of \bJ1653.
\label{table-all1653}
\end{sidewaystable}

\addtocounter{table}{-1}
\begin{sidewaystable}[ht]
\footnotesize
\begin{center}
\caption{({\it continued})}
    \tabcolsep 4.3pt
\begin{tabular}{lcccccccccccccccc}
\hline\hline
\multicolumn{1}{c}{N} &
\multicolumn{1}{c}{CXOU J} &
\multicolumn{1}{c}{R.A.} &
\multicolumn{1}{c}{Decl.} &
\multicolumn{1}{c}{$r$} &
\multicolumn{1}{c}{$D$} &
\multicolumn{1}{c}{$A_{\rm eff}$} &
\multicolumn{1}{c}{log} &
\multicolumn{1}{c}{Counts} &
\multicolumn{1}{c}{$\sigma$} &
\multicolumn{1}{c}{$P_{\rm B}$} &
\multicolumn{1}{c}{Counts} &
\multicolumn{1}{c}{$\sigma$} &
\multicolumn{1}{c}{$P_{\rm B}$} &
\multicolumn{1}{c}{Counts} &
\multicolumn{1}{c}{$\sigma$} &
\multicolumn{1}{c}{$P_{\rm B}$} \\
\cline{9-11}\cline{12-14}\cline{15-17}
\multicolumn{1}{c}{} &
\multicolumn{1}{c}{Name} &
\multicolumn{1}{c}{(deg)} &
\multicolumn{1}{c}{(deg)} &
\multicolumn{1}{c}{($\arcsec$)} &
\multicolumn{1}{c}{($'$)} &
\multicolumn{1}{c}{(cm$^2$)} &
\multicolumn{1}{c}{$P_{\rm KS}$} &
\multicolumn{3}{c}{($0.5-8$ keV)} &
\multicolumn{3}{c}{($0.5-2$ keV)} &
\multicolumn{3}{c}{($2-8$ keV)} \\
\hline
 51 & 165343.4-015841 & 253.43123 &  -1.97813 & 0.16 & 0.7 & 500 & -0.43 &    5.9 +3.6/-2.4 & 1.6 & 8.2e-11 &    6.0 +3.6/-2.4 & 1.7 & 1.3e-13 &   0.0 +1.9/0.0 & 0.0 & 1.0 \\ 
 52 & 165344.0-014945 & 253.43351 &  -1.82921 & 1.23 & 8.3 & 355 & -0.21 &    7.0 +4.3/-3.1 & 1.6 & 1.5e-03 &    3.0 +3.2/-1.9 & 0.9 & 2.4e-02 &    4.0 +3.6/-2.4 & 1.1 & 1.9e-02 \\ 
 53 & 165344.1-020355 & 253.43404 &  -2.06545 & 1.17 & 5.9 & 438 & .. &    2.3 +2.9/-1.6 & 0.8 & 3.5e-02 &   -0.3 +1.9/0.0 & -0.2 & 1.0 &    2.6 +2.9/-1.6 & 0.9 & 8.4e-03 \\ 
 54 & 165344.4-015306 & 253.43509 &  -1.88505 & 0.67 & 4.9 & 462 & -0.13 &    3.5 +3.2/-1.9 & 1.1 & 1.5e-03 &   -0.2 +1.9/0.0 & -0.1 & 1.0 &    3.7 +3.2/-1.9 & 1.2 & 1.8e-04 \\ 
 55 & 165345.4-020434 & 253.43935 &  -2.07631 & 1.12 & 6.6 & 445 & -0.40 &    3.8 +3.4/-2.2 & 1.1 & 7.9e-03 &   -0.4 +1.9/0.0 & -0.2 & 1.0 &    4.3 +3.4/-2.2 & 1.2 & 1.2e-03 \\ 
 56 & 165345.5-020209 & 253.43993 &  -2.03600 & 0.63 & 4.2 & 494 & .. &    2.7 +2.9/-1.6 & 0.9 & 5.1e-03 &    0.8 +2.3/-0.8 & 0.4 & 1.5e-01 &    1.8 +2.7/-1.3 & 0.7 & 1.5e-02 \\ 
 57 & 165346.0-020133 & 253.44181 &  -2.02587 & 0.28 & 3.6 & 495 & -0.08 &    8.8 +4.1/-2.9 & 2.1 & 1.6e-12 &    6.9 +3.8/-2.6 & 1.8 & 3.9e-12 &    1.9 +2.7/-1.3 & 0.7 & 7.6e-03 \\ 
 58 & 165346.7-020109 & 253.44466 &  -2.01930 & 0.21 & 3.2 & 501 & -0.03 &   11.8 +4.6/-3.4 & 2.6 & 2.7e-18 &    5.9 +3.6/-2.4 & 1.6 & 6.1e-11 &    5.9 +3.6/-2.4 & 1.6 & 4.8e-09 \\ 
 59 & 165347.0-015403 & 253.44594 &  -1.90108 & 0.27 & 4.0 & 444 & -0.25 &   11.8 +4.6/-3.4 & 2.6 & 1.7e-16 &   10.9 +4.4/-3.3 & 2.5 & 7.7e-19 &    0.9 +2.3/-0.8 & 0.4 & 1.4e-01 \\ 
 60 & 165347.0-015054 & 253.44604 &  -1.84837 & 0.43 & 7.2 & 437 & -0.07 &   35.2 +7.1/-6.1 & 4.9 & 8.9e-33 &   23.4 +6.0/-4.9 & 3.9 & 3.2e-27 &   11.8 +4.7/-3.6 & 2.5 & 1.7e-09 \\ 
 61 & 165347.7-020610 & 253.44878 &  -2.10295 & 1.89 & 8.2 & 400 & -0.32 &    2.2 +3.4/-2.2 & 0.7 & 1.5e-01 &    4.0 +3.4/-2.2 & 1.2 & 5.4e-03 &   -1.7 +1.9/0.0 & -0.9 & 1.0 \\ 
 62 & 165348.0-015400 & 253.45007 &  -1.90003 & 0.15 & 4.1 & 347 & -0.09 &   42.8 +7.6/-6.5 & 5.6 & 0.0 &   19.9 +5.6/-4.4 & 3.6 & 0.0 &   22.9 +5.9/-4.8 & 3.9 & 0.0 \\ 
 63 & 165348.8-015413 & 253.45368 &  -1.90380 & 0.50 & 4.0 & 435 & -0.30 &    3.8 +3.2/-1.9 & 1.2 & 1.3e-04 &    2.9 +2.9/-1.6 & 1.0 & 1.4e-04 &    0.8 +2.3/-0.8 & 0.4 & 1.4e-01 \\ 
 64 & 165349.4-015818 & 253.45593 &  -1.97174 & 0.20 & 1.4 & 310 & -0.19 &    6.0 +3.6/-2.4 & 1.7 & 3.1e-11 &    1.0 +2.3/-0.8 & 0.4 & 1.1e-02 &    5.0 +3.4/-2.2 & 1.5 & 1.5e-09 \\ 
 65 & 165349.6-015414 & 253.45701 &  -1.90397 & 0.60 & 4.0 & 476 & .. &    2.7 +2.9/-1.6 & 0.9 & 2.6e-03 &    1.9 +2.7/-1.3 & 0.7 & 5.5e-03 &    0.8 +2.3/-0.8 & 0.4 & 1.5e-01 \\ 
 66 & 165349.9-015242 & 253.45828 &  -1.87845 & 0.89 & 5.5 & 309 & .. &    2.4 +2.9/-1.6 & 0.8 & 2.0e-02 &    1.7 +2.7/-1.3 & 0.7 & 3.0e-02 &    0.7 +2.3/-0.8 & 0.3 & 2.6e-01 \\ 
 67 & 165350.4-015508 & 253.46021 &  -1.91900 & 0.15 & 3.3 & 456 & -0.40 &   26.9 +6.3/-5.2 & 4.3 & 0.0 &    8.0 +4.0/-2.8 & 2.0 & 6.7e-16 &   18.9 +5.4/-4.3 & 3.5 & 8.9e-37 \\ 
 68 & 165350.5-015825 & 253.46055 &  -1.97379 & 0.30 & 1.7 & 301 & .. &    3.0 +2.9/-1.6 & 1.0 & 1.6e-05 &   0.0 +1.9/0.0 & 0.0 & 1.0 &    3.0 +2.9/-1.6 & 1.0 & 3.9e-06 \\ 
 69 & 165351.9-015847 & 253.46666 &  -1.97973 & 0.20 & 2.2 & 214 & -0.17 &    7.0 +3.8/-2.6 & 1.8 & 4.7e-14 &    3.0 +2.9/-1.6 & 1.0 & 1.2e-06 &    4.0 +3.2/-1.9 & 1.2 & 1.2e-08 \\ 
 70 & 165356.0-020251 & 253.48362 &  -2.04764 & 0.37 & 5.7 & 477 & -1.68 &   21.0 +5.8/-4.7 & 3.6 & 4.2e-21 &   12.6 +4.7/-3.6 & 2.7 & 4.2e-15 &    8.4 +4.1/-2.9 & 2.0 & 4.2e-08 \\ 
 71 & 165357.4-015239 & 253.48957 &  -1.87760 & 0.56 & 6.4 & 447 & -0.15 &   14.5 +5.1/-4.0 & 2.8 & 2.4e-11 &    6.4 +3.8/-2.6 & 1.7 & 5.0e-06 &    8.1 +4.1/-2.9 & 2.0 & 1.1e-06 \\ 
 72 & 165357.6-015740 & 253.49014 &  -1.96121 & 0.27 & 3.5 & 487 & -0.10 &    8.8 +4.1/-2.9 & 2.1 & 5.4e-13 &    6.9 +3.8/-2.6 & 1.8 & 4.5e-12 &    1.9 +2.7/-1.3 & 0.7 & 4.9e-03 \\ 
 73 & 165358.4-015526 & 253.49364 &  -1.92406 & 0.60 & 4.5 & 481 & -0.10 &    3.6 +3.2/-1.9 & 1.1 & 6.4e-04 &   -0.1 +1.9/0.0 & -0.1 & 1.0 &    3.8 +3.2/-1.9 & 1.2 & 1.2e-04 \\ 
 74 & 165358.6-020018 & 253.49433 &  -2.00525 & 0.69 & 4.4 & 342 & .. &    2.7 +2.9/-1.6 & 0.9 & 3.3e-03 &    0.9 +2.3/-0.8 & 0.4 & 9.8e-02 &    1.8 +2.7/-1.3 & 0.7 & 1.6e-02 \\ 
 75 & 165359.0-020316 & 253.49594 &  -2.05447 & 0.12 & 6.5 & 385 & -0.08 &  312.5 +18.7/-17.7 & 16.7 & 0.0 &  170.4 +14.1/-13.1 & 12.1 & 0.0 &  142.1 +13.0/-11.9 & 10.9 & 0.0 \\ 
 76 & 165359.3-015630 & 253.49718 &  -1.94175 & 0.60 & 4.2 & 451 & .. &    2.7 +2.9/-1.6 & 0.9 & 3.1e-03 &    1.9 +2.7/-1.3 & 0.7 & 5.2e-03 &    0.8 +2.3/-0.8 & 0.3 & 1.6e-01 \\ 
 77 & 165400.4-015454 & 253.50192 &  -1.91503 & 0.90 & 5.2 & 469 & .. &    2.5 +2.9/-1.6 & 0.8 & 1.9e-02 &   -0.2 +1.9/0.0 & -0.1 & 1.0 &    2.7 +2.9/-1.6 & 0.9 & 4.8e-03 \\ 
 78 & 165400.5-015826 & 253.50237 &  -1.97406 & 0.67 & 4.2 & 448 & .. &    2.6 +2.9/-1.6 & 0.9 & 7.9e-03 &    1.9 +2.7/-1.3 & 0.7 & 1.0e-02 &    0.8 +2.3/-0.8 & 0.3 & 2.2e-01 \\ 
 79 & 165400.8-015549 & 253.50362 &  -1.93053 & 0.77 & 4.8 & 476 & .. &    2.6 +2.9/-1.6 & 0.9 & 1.0e-02 &   -0.2 +1.9/0.0 & -0.1 & 1.0 &    2.7 +2.9/-1.6 & 0.9 & 2.6e-03 \\ 
 80 & 165401.1-020048 & 253.50496 &  -2.01335 & 0.31 & 5.2 & 432 & -1.11 &   24.3 +6.1/-5.0 & 4.0 & 1.4e-29 &   21.7 +5.8/-4.7 & 3.8 & 1.3e-32 &    2.6 +2.9/-1.6 & 0.9 & 8.5e-03 \\ 
 81 & 165401.5-015053 & 253.50641 &  -1.84824 & 1.19 & 8.4 & 338 & -1.82 &    9.1 +4.6/-3.4 & 2.0 & 8.3e-05 &    4.1 +3.4/-2.2 & 1.2 & 3.4e-03 &    5.0 +3.8/-2.6 & 1.3 & 5.5e-03 \\ 
 82 & 165402.0-015854 & 253.50846 &  -1.98171 & 0.43 & 4.6 & 439 & -0.02 &    8.6 +4.1/-2.9 & 2.1 & 5.1e-10 &    4.9 +3.4/-2.2 & 1.4 & 4.7e-07 &    3.8 +3.2/-1.9 & 1.2 & 1.5e-04 \\ 
 83 & 165402.1-015322 & 253.50909 &  -1.88970 & 1.12 & 6.5 & 455 & -0.28 &    3.6 +3.4/-2.2 & 1.1 & 1.7e-02 &    4.4 +3.4/-2.2 & 1.3 & 4.5e-04 &   -0.9 +1.9/0.0 & -0.5 & 1.0 \\ 
 84 & 165402.9-015425 & 253.51220 &  -1.90706 & 0.69 & 6.0 & 461 & -0.16 &    7.0 +4.0/-2.8 & 1.8 & 9.6e-06 &    4.6 +3.4/-2.2 & 1.4 & 5.4e-05 &    2.4 +2.9/-1.6 & 0.8 & 2.4e-02 \\ 
 85 & 165404.5-020346 & 253.51891 &  -2.06305 & 0.59 & 7.7 & 423 & -0.49 &   26.0 +6.5/-5.4 & 4.0 & 1.1e-17 &   18.9 +5.6/-4.4 & 3.4 & 8.6e-17 &    7.1 +4.1/-2.9 & 1.7 & 2.5e-04 \\ 
 86 & 165405.6-015905 & 253.52360 &  -1.98490 & 0.60 & 5.6 & 473 & -0.13 &    9.1 +4.3/-3.1 & 2.1 & 4.7e-08 &   -0.3 +1.9/0.0 & -0.2 & 1.0 &    9.4 +4.3/-3.1 & 2.2 & 9.1e-10 \\ 
 87 & 165406.9-020540 & 253.52901 &  -2.09457 & 0.90 & 9.6 & 423 & -0.12 &   22.9 +6.5/-5.4 & 3.5 & 6.9e-11 &   12.5 +5.0/-3.8 & 2.5 & 2.3e-07 &   10.5 +4.9/-3.7 & 2.2 & 3.6e-05 \\ 
 88 & 165407.3-015957 & 253.53044 &  -1.99918 & 1.19 & 6.2 & 463 & -0.35 &    2.8 +3.2/-1.9 & 0.9 & 3.9e-02 &    3.5 +3.2/-1.9 & 1.1 & 2.0e-03 &   -0.7 +1.9/0.0 & -0.4 & 1.0 \\ 
 89 & 165409.8-020219 & 253.54089 &  -2.03870 & 0.54 & 7.8 & 445 & -0.16 &   32.0 +7.0/-5.9 & 4.6 & 7.5e-24 &   18.8 +5.6/-4.4 & 3.4 & 9.2e-17 &   13.2 +5.0/-3.8 & 2.7 & 2.6e-09 \\ 
 90 & 165409.9-015736 & 253.54128 &  -1.96018 & 1.44 & 6.5 & 456 & -0.55 &    2.8 +3.2/-1.9 & 0.9 & 3.3e-02 &    3.6 +3.2/-1.9 & 1.1 & 1.0e-03 &   -0.8 +1.9/0.0 & -0.4 & 1.0 \\ 
 91 & 165411.9-015610 & 253.54964 &  -1.93620 & 0.72 & 7.3 & 450 & -0.03 &   12.8 +5.0/-3.8 & 2.6 & 2.7e-08 &   10.2 +4.4/-3.3 & 2.3 & 3.5e-09 &    2.6 +3.2/-1.9 & 0.8 & 5.7e-02 \\ 
 92 & 165412.6-015701 & 253.55287 &  -1.95052 & 0.81 & 7.3 & 452 & -1.72 &   10.1 +4.6/-3.4 & 2.2 & 1.7e-06 &    4.4 +3.4/-2.2 & 1.3 & 6.5e-04 &    5.7 +3.8/-2.6 & 1.5 & 5.7e-04 \\ 
 93 & 165414.5-015707 & 253.56049 &  -1.95214 & 1.38 & 7.7 & 395 & -0.10 &    4.9 +3.8/-2.6 & 1.3 & 6.5e-03 &   -0.6 +1.9/0.0 & -0.3 & 1.0 &    5.5 +3.8/-2.6 & 1.4 & 1.2e-03 \\ 
 94 & 165414.8-015442 & 253.56181 &  -1.91192 & 0.69 & 8.4 & 435 & -0.33 &   24.7 +6.4/-5.3 & 3.9 & 6.5e-16 &   16.0 +5.2/-4.1 & 3.1 & 1.9e-14 &    8.6 +4.4/-3.3 & 1.9 & 6.0e-05 \\ 
 95 & 165415.3-015455 & 253.56394 &  -1.91547 & 1.79 & 8.5 & 435 & -0.25 &    2.4 +3.6/-2.4 & 0.7 & 1.7e-01 &    2.7 +3.2/-1.9 & 0.8 & 4.8e-02 &   -0.3 +2.7/-1.3 & -0.1 & 6.8e-01 \\ 
 96 & 165416.8-015749 & 253.57027 &  -1.96382 & 1.45 & 8.3 & 437 & -0.09 &    4.6 +4.0/-2.8 & 1.2 & 2.6e-02 &    5.0 +3.6/-2.4 & 1.4 & 1.1e-03 &   -0.4 +2.7/-1.3 & -0.1 & 6.9e-01 \\ 
 97 & 165418.2-020018 & 253.57588 &  -2.00523 & 1.30 & 8.9 & 363 & -0.54 &    7.7 +4.6/-3.4 & 1.7 & 2.0e-03 &    6.3 +4.0/-2.8 & 1.6 & 5.3e-04 &    1.4 +3.2/-1.9 & 0.4 & 2.7e-01 \\ 
\hline
\hline
\end{tabular}
\end{center}
\end{sidewaystable}

\clearpage
\newpage

\section{B. Optical, Near-infrared, and Mid-Infrared Counterparts 
\label{sec-optical}}

We cross-matched the \Chandra\ X-ray source lists (Appendix~A) for 
\aJ1311\ and \bJ1653\ with the USNO~B1.0 optical, 2MASS near-infrared, 
and WISEP mid-infrared catalogs. The WISE bands are in the 3.4 ($W1$), 
4.6 ($W2$), 12 ($W3$), and 22$\mu$m ($W4$) bandpasses. In order to 
include as many potential matches as practical, we used a relatively 
liberal search radius of 3\arcsec\ from the \Chandra\ positions, 
although a comparison of the \Chandra\ images with the optical/IR ones 
show the most believable counterparts have $< 1\arcsec - 2\arcsec$ 
offsets. Out of the 97 sources found in each of the \Chandra\ fields, we 
located potential USNO~B1.0 counterparts for up to 17 (one with two 
optical matches) and 16 (four with two optical matches) X-ray sources in 
the \aJ1311\ and \bJ1653\ fields, respectively (Table~\ref{table-usno}). 
In the 2MASS catalog, we found up to six and eight matches for these 
respective fields (Table~\ref{table-2mass}). In the WISEP catalog, there 
were up to 25 and 21 matches, respectively (Table~\ref{table-wise}). All 
2MASS matches were also found in the USNO~B1.0 and WISEP catalogs. In 
the \aJ1311\ and \bJ1653\ fields, seven and three WISEP sources 
respectively, had USNO~B1.0 counterparts but no 2MASS ones. There were 
12 and 10 WISEP matches with X-ray sources without matches in the 
USNO~B1.0 and 2MASS catalogs in the two respective fields.

To quantify possible systematic errors in the \Chandra\ positions, we 
calculated the differences, $\delta$(R.A., Decl.) = (R.A., Decl.)$_{\rm 
USNO}$ $-$ (R.A., Decl.)$_{Chandra}$, for all X-ray sources with $\simgt 
4$ net counts and detected at $>1.0 \sigma$ that are within 1\arcsec\ of 
an USNO source. For 11 sources in the \aJ1311\ and 10 sources in the 
\bJ1653\ fields, we found average offsets, $\delta$(R.A., Decl.) = 
$(-0.06\arcsec \pm 0.44\arcsec, -0.08\arcsec \pm 0.44\arcsec)$ and 
$(0.36\arcsec \pm 0.34\arcsec, 0.37\arcsec \pm 0.18\arcsec)$, 
respectively. Thus, only the X-ray positions in the \bJ1653\ field 
appear to have a significant systematic offset amounting to $r_{\rm 
syst.} = 0.52\arcsec \pm 0.39\arcsec$.

\begin{table}[b]
\footnotesize
\begin{center}
\caption{}
\begin{tabular}{lccccccc}
\hline\hline
\multicolumn{1}{c}{N} &
\multicolumn{1}{c}{USNO-B1.0 name} &
\multicolumn{1}{c}{offset} &
\multicolumn{1}{c}{$B2$} &
\multicolumn{1}{c}{$R2$} &
\multicolumn{1}{c}{$I$} &
\multicolumn{2}{c}{USNO~B1.0 position} \\
\cline{7-8}
\multicolumn{1}{c}{} &
\multicolumn{1}{c}{} &
\multicolumn{1}{c}{(\arcsec)} &
\multicolumn{1}{c}{(mag)} &
\multicolumn{1}{c}{(mag)} &
\multicolumn{1}{c}{(mag)} &
\multicolumn{1}{c}{R.A.} &
\multicolumn{1}{c}{Decl.} \\
\hline
\multicolumn{8}{c}{\aJ1311} \\
\hline
20 & 0555-0290744 & 1.06 & 16.07 & 14.89 & 14.32 & 13 11 34.425 & -34 29 20.44 \\ 
31 & 0555-0290806 & 0.37 & 19.67 & 18.87 & 18.33 & 13 11 43.735 & -34 27 49.78 \\ 
35 & 0554-0289419 & 0.62 & 21.02 & .. & .. & 13 11 45.741 & -34 30 29.96 \\ 
43 & 0553-0287487 & 0.58 & 20.09 & 20.03 & .. & 13 11 49.204 & -34 38 17.31 \\ 
45 & 0554-0289445 & 1.67 & 17.53 & 15.65 & 14.92 & 13 11 49.518 & -34 33 31.73 \\ 
54 & 0555-0290875 & 0.25 & 18.03 & 16.90 & 16.72 & 13 11 54.978 & -34 26 42.54 \\ 
61 & 0555-0290895 & 0.62 & 20.88 & 19.97 & .. & 13 11 58.216 & -34 27 33.73 \\ 
.. & 0555-0290894 & 1.36 & .. & 19.15 & 18.49 & 13 11 58.165 & -34 27 34.77 \\ 
64 & 0554-0289501 & 0.66 & 20.80 & 19.89 & .. & 13 11 59.537 & -34 35 54.96 \\ 
66 & 0554-0289512 & 0.74 & 16.63 & 15.50 & 14.22 & 13 12 02.102 & -34 32 34.62 \\ 
69 & 0556-0290675 & 2.30 & 16.67 & 14.48 & 13.05 & 13 12 04.097 & -34 20 26.25 \\ 
75 & 0554-0289533 & 0.34 & 21.00 & 20.39 & .. & 13 12 06.126 & -34 31 52.63 \\ 
78 & 0554-0289544 & 0.22 & 21.19 & .. & .. & 13 12 07.411 & -34 31 07.71 \\ 
82 & 0554-0289558 & 1.81 & 18.75 & 18.15 & 17.53 & 13 12 09.823 & -34 34 38.80 \\ 
85 & 0555-0290992 & 0.95 & 21.08 & .. & 18.69 & 13 12 18.364 & -34 28 19.28 \\ 
89 & 0555-0291020 & 0.88 & 18.97 & 18.69 & 17.82 & 13 12 23.109 & -34 29 12.04 \\ 
91 & 0555-0291038 & 0.20 & 13.11 & 11.87 & 11.38 & 13 12 25.362 & -34 25 01.51 \\ 
96 & 0555-0291083 & 1.02 & 20.48 & 19.88 & .. & 13 12 34.909 & -34 29 25.27 \\ 
\hline
\hline
\multicolumn{8}{c}{\bJ1653} \\
\hline
8 & 0880-0368721 & 0.50 & 18.47 & 16.97 & 16.96 & 16 53 15.616 & -01 58 22.66 \\ 
.. & 0880-0368718 & 1.85 & .. & 18.01 & 19.27 & 16 53 15.537 & -01 58 23.75 \\ 
23 & 0881-0336447 & 2.75 & 20.59 & 19.45 & 20.06 & 16 53 27.505 & -01 51 15.69 \\ 
25 & 0880-0368904 & 1.62 & .. & 19.89 & .. & 16 53 29.935 & -01 58 28.39 \\
.. & 0880-0368902 & 1.67 & 20.47 & 19.28 & 18.13 & 16 53 29.928 & -01 58 27.65 \\ 
28 & 0881-0336518 & 0.79 & 10.78 & 9.56 & 9.05 & 16 53 33.598 & -01 52 59.72 \\ 
30 & 0879-0416563 & 1.77 & .. & 20.39 & 19.33 & 16 53 34.735 & -02 00 31.96 \\ 
.. & 0879-0416560 & 2.95 & .. & 19.43 & 19.10 & 16 53 34.686 & -02 00 30.07 \\ 
36 & 0879-0416588 & 0.53 & 19.72 & 17.93 & 18.01 & 16 53 37.207 & -02 04 42.56 \\ 
38 & 0880-0369025 & 0.28 & 20.40 & 19.41 & 20.0 & 16 53 38.070 & -01 58 36.71 \\ 
40 & 0880-0369044 & 0.58 & 20.64 & 19.42 & 18.54 & 16 53 39.115 & -01 54 12.08 \\ 
44 & 0880-0369077 & 0.68 & 20.58 & 18.95 & 17.05 & 16 53 41.410 & -01 59 27.68 \\ 
.. & 0880-0369078 & 2.36 & 20.89 & 19.29 & .. & 16 53 41.438 & -01 59 29.31 \\ 
49 & 0879-0416658 & 0.32 & 16.64 & 14.60 & 14.16 & 16 53 42.849 & -02 01 45.09 \\ 
59 & 0880-0369161 & 0.73 & 19.05 & 16.05 & 15.06 & 16 53 47.071 & -01 54 04.14 \\ 
60 & 0881-0336669 & 0.37 & 20.35 & 19.38 & .. & 16 53 47.030 & -01 50 54.32 \\ 
61 & 0878-0449662 & 2.87 & .. & 19.44 & 19.44 & 16 53 47.630 & -02 06 13.25 \\ 
75 & 0879-0416898 & 2.27 & 18.75 & 17.31 & 16.49 & 16 53 59.085 & -02 03 13.99 \\ 
80 & 0879-0416929 & 0.72 & 20.58 & 18.62 & 17.01 & 16 54 01.230 & -02 00 48.46 \\ 
97 & 0879-0417221 & 0.54 & 13.94 & 12.27 & 12.02 & 16 54 18.222 & -02 00 19.33 \\ 
\hline
\hline
\end{tabular}
\end{center}
Potential counterpart matches to the \Chandra\ X-ray sources from the USNO-B1.0 catalog.
\label{table-usno}
\end{table}

\begin{table}
\footnotesize
\begin{center}
\caption{}
\begin{tabular}{lccccc}
\hline\hline
\multicolumn{1}{c}{N} &
\multicolumn{1}{c}{2MASS name} &
\multicolumn{1}{c}{offset} &
\multicolumn{1}{c}{$J$ (error)} &
\multicolumn{1}{c}{$H$ (error)} &
\multicolumn{1}{c}{$K$ (error)} \\
\multicolumn{1}{c}{} &
\multicolumn{1}{c}{} &
\multicolumn{1}{c}{(\arcsec)} &
\multicolumn{1}{c}{(mag)} &
\multicolumn{1}{c}{(mag)} &
\multicolumn{1}{c}{(mag)} \\
\hline
\multicolumn{6}{c}{\aJ1311} \\
\hline
20 & J13113441-3429208 & 0.64 & 14.324 (0.027) & 14.038 (0.038) & 13.998 (0.045) \\
45 & J13114950-3433318 & 1.93 & 14.449 (0.032) & 13.934 (0.027) & 13.834 (0.037) \\
54 & J13115497-3426426 & 0.16 & 16.099 (0.100) & 15.444 (0.086) & 14.999 (0.124) \\
66 & J13120210-3432341 & 0.61 & 14.588 (..) & 14.629 (0.080) & 13.613 (..) \\
69 & J13120408-3420263 & 2.19 & 12.636 (0.026) & 11.960 (0.025) & 11.825 (0.023) \\
91 & J13122539-3425015 & 0.42 & 11.294 (0.023) & 10.971 (0.023) & 10.892 (0.023) \\
\hline
\hline
\multicolumn{6}{c}{\bJ1653} \\
\hline
8 & J16531562-0158224 & 0.35 & 16.196 (0.127) & 15.560 (0.147) & 14.292 (0.088) \\
28 & J16533360-0152597 & 0.89 & 8.522 (0.020) & 8.067 (0.020) & 7.996 (0.023) \\
44 & J16534140-0159272 & 0.27 & 15.094 (0.056) & 14.469 (0.059) & 14.147 (0.066) \\
49 & J16534285-0201450 & 0.31 & 13.133 (0.023) & 12.575 (0.026) & 12.492 (0.030) \\
59 & J16534707-0154041 & 0.73 & 13.634 (0.033) & 12.942 (0.034) & 12.737 (0.036) \\
75 & J16535902-0203163 & 0.30 & 16.488 (0.195) & 15.624 (0.153) & 14.575 (0.112) \\
80 & J16540121-0200480 & 0.31 & 15.007 (0.057) & 14.380 (0.055) & 14.043 (0.076) \\
97 & J16541823-0200193 & 0.59 & 11.079 (0.021) & 10.696 (0.023) & 10.539 (0.022) \\
\hline
\hline
\end{tabular}
\end{center}
Potential counterpart matches to the \Chandra\ X-ray sources in the 2MASS catalog.
\label{table-2mass}
\end{table}

\begin{table}[ht]
\footnotesize
\begin{center}
\caption{}
\begin{tabular}{lcccccc}
\hline\hline
\multicolumn{1}{c}{N} &
\multicolumn{1}{c}{WISEP name} &
\multicolumn{1}{c}{offset} &
\multicolumn{1}{c}{$W1$ (error)} &
\multicolumn{1}{c}{$W2$ (error)} &
\multicolumn{1}{c}{$W3$ (error)} &
\multicolumn{1}{c}{$W4$ (error)} \\
\multicolumn{1}{c}{} &
\multicolumn{1}{c}{} &
\multicolumn{1}{c}{(\arcsec)} &
\multicolumn{1}{c}{(mag)} &
\multicolumn{1}{c}{(mag)} &
\multicolumn{1}{c}{(mag)} &
\multicolumn{1}{c}{(mag)} \\
\hline
\multicolumn{7}{c}{\aJ1311} \\
\hline
13 & J131129.38-343134.0 & 0.77 & 16.899 (0.150) & 15.992 (0.232) & 12.732 (..) & 8.665 (..) \\ 
20 & J131134.41-342920.6 & 0.78 & 13.853 (0.031) & 13.840 (0.047) & 12.449 (..) & 8.534 (..) \\ 
29 & J131142.16-342955.8 & 2.56 & 16.397 (0.101) & 16.042 (0.233) & 12.343 (..) & 9.180 (..) \\ 
31 & J131143.72-342749.9 & 0.18 & 15.790 (0.067) & 15.302 (0.128) & 11.432 (0.169) & 7.954 (0.193) \\ 
32 & J131143.98-342500.4 & 0.28 & 15.518 (0.057) & 15.039 (0.111) & 12.290 (0.364) & 9.219 (..) \\ 
34 & J131145.72-343312.4 & 1.50 & 16.197 (0.080) & 15.527 (0.150) & 12.766 (..) & 9.008 (..) \\ 
39 & J131147.09-343205.3 & 0.69 & 15.720 (0.063) & 14.949 (0.095) & 12.226 (0.333) & 9.056 (0.463) \\ 
40 & J131147.15-342953.5 & 1.25 & 16.846 (0.152) & 16.408 (0.345) & 12.443 (..) & 8.721 (..) \\ 
43 & J131149.24-343817.3 & 0.30 & 16.639 (0.135) & 15.609 (0.177) & 12.457 (0.450) & 8.850 (..) \\ 
45 & J131149.51-343331.4 & 1.57 & 13.762 (0.030) & 13.750 (0.043) & 12.498 (..) & 9.219 (..) \\ 
54 & J131154.98-342642.1 & 0.63 & 14.474 (0.034) & 14.207 (0.060) & 12.009 (..) & 8.806 (..) \\ 
55 & J131156.02-342124.8 & 2.32 & 16.738 (0.149) & 16.522 (0.405) & 12.550 (..) & 8.838 (..) \\ 
64 & J131159.49-343553.6 & 2.03 & 16.539 (0.115) & 15.237 (0.121) & 12.548 (0.448) & 8.655 (..) \\ 
65 & J131200.45-343231.4 & 0.56 & 16.542 (0.115) & 15.604 (0.168) & 12.172 (..) & 9.120 (..) \\ 
66 & J131202.10-343235.1 & 1.19 & 13.230 (0.027) & 13.058 (0.033) & 10.312 (0.069) & 7.975 (0.197) \\ 
69 & J131204.09-342026.3 & 2.27 & 11.709 (0.025) & 11.669 (0.023) & 11.842 (0.234) & 9.138 (..) \\ 
70 & J131203.88-343652.3 & 1.11 & 17.032 (0.187) & 15.248 (0.125) & 12.540 (0.480) & 9.006 (..) \\ 
75 & J131206.11-343151.3 & 1.15 & 16.081 (0.107) & 15.533 (0.173) & 12.090 (..) & 8.641 (..) \\ 
76 & J131206.16-343446.8 & 2.86 & 13.466 (0.027) & 13.034 (0.034) & 11.171 (0.134) & 8.687 (0.379) \\ 
77 & J131206.59-342909.6 & 0.76 & 16.366 (0.109) & 16.211 (0.303) & 12.652 (..) & 8.773h (..) \\ 
78 & J131207.47-343107.8 & 0.89 & 16.417 (0.112) & 16.119 (0.291) & 12.307 (..) & 8.833 (..) \\ 
82 & J131209.82-343438.9 & 1.67 & 15.317 (0.050) & 15.059 (0.106) & 12.203 (..) & 8.669 (0.336) \\ 
87 & J131220.24-342930.7 & 0.94 & 16.492 (0.122) & 16.160 (0.281) & 12.384 (0.416) & 8.948 (..) \\ 
89 & J131223.15-342912.2 & 0.57 & 15.169 (0.049) & 14.785 (0.087) & 11.711 (0.228) & 8.504 (..) \\ 
91 & J131225.40-342501.5 & 0.50 & 10.817 (0.024) & 10.806 (0.022) & 10.612 (0.086) & 8.546 (..) \\ 
\hline
\hline
\multicolumn{7}{c}{\bJ1653} \\
\hline
2 & J165307.76-015955.5 & 1.46 & 15.841 (0.069) & 14.383 (0.066) & 11.470 (0.180) & 8.545 (0.373) \\ 
3 & J165309.03-020139.0 & 0.70 & 15.791 (0.064) & 15.597 (0.163) & 12.478 (..) & 9.131 (..) \\ 
4 & J165309.82-015915.2 & 1.70 & 15.608 (0.060) & 15.257 (0.128) & 12.032 (..) & 9.141 (..) \\ 
7 & J165315.14-020314.2 & 1.48 & 16.552 (0.113) & 16.642 (0.403) & 12.076 (..) & 8.810 (..) \\ 
8 & J165315.62-015822.3 & 0.41 & 12.599 (0.025) & 11.578 (0.025) & 9.208 (0.037) & 7.364 (0.113) \\ 
11 & J165318.45-015432.3 & 2.45 & 16.885 (0.154) & 16.508 (0.376) & 12.306 (0.388) & 8.957 (..) \\ 
14 & J165323.24-020451.1 & 0.34 & 16.240 (0.089) & 15.220 (0.121) & 12.150 (..) & 9.074 (..) \\ 
25 & J165329.94-015826.7 & 1.78 & 15.589 (0.057) & 15.529 (0.159) & 12.340 (..) & 9.112 (..) \\ 
28 & J165333.57-015259.6 & 0.38 & 7.934 (0.021) & 7.990 (0.023) & 7.915 (0.023) & 7.926 (0.198) \\ 
36 & J165337.19-020442.5 & 0.41 & 16.385 (0.093) & 16.687 (0.417) & 12.783 (..) & 9.083 (..) \\ 
44 & J165341.41-015927.6 & 0.65 & 14.057 (0.031) & 13.843 (0.046) & 12.569 (..) & 8.667 (..) \\ 
49 & J165342.83-020145.1 & 0.25 & 12.420 (0.026) & 12.429 (0.027) & 12.292 (0.413) & 9.014 (..) \\ 
51 & J165343.52-015840.6 & 0.69 & 16.245 (0.100) & 16.366 (0.352) & 12.612 (..) & 8.882 (..) \\ 
59 & J165347.05-015404.4 & 0.77 & 12.649 (0.025) & 12.584 (0.029) & 12.148 (..) & 8.501 (..) \\ 
60 & J165347.06-015053.8 & 0.35 & 17.021 (0.174) & 15.530 (0.153) & 12.586 (0.493) & 8.923 (..) \\ 
67 & J165350.45-015509.2 & 0.86 & 15.655 (0.066) & 14.910 (0.098) & 12.186 (0.365) & 8.562 (..) \\ 
75 & J165359.08-020313.9 & 2.29 & 12.765 (0.025) & 12.142 (0.025) & 9.840 (0.050) & 7.213 (0.095) \\ 
80 & J165401.19-020048.3 & 0.34 & 14.007 (0.030) & 13.776 (0.045) & 11.992 (..) & 9.009 (..) \\ 
86 & J165405.67-015905.1 & 0.57 & 15.748 (0.061) & 14.702 (0.083) & 11.254 (0.155) & 8.766 (0.431) \\ 
87 & J165406.91-020540.1 & 0.73 & 17.036 (0.182) & 15.394 (0.140) & 11.672 (0.208) & 8.533 (..) \\ 
97 & J165418.21-020019.3 & 0.49 & 10.490 (0.023) & 10.515 (0.021) & 10.583 (0.088) & 8.850 (..) \\ 
\hline
\hline
\end{tabular}
\end{center}
Potential counterpart matches to the \Chandra\ X-ray sources in the WISEP catalog. A single source is flagged with an ``h'' in the $W4$ band 
(J131206.59$-$342909.6, N77) as being possibly spurious or with photometry contaminated by scattered light from a nearby bright source.
\label{table-wise}
\end{table}


\begin{thebibliography}{}

\bibitem[Abdo et al.(2009a)]{bsl} Abdo, A.~A., et al.\ (\Fermi-LAT 
collaboration) 2009a, \apjs, 183, 46 (LAT Bright Source List)

\bibitem[Abdo et al.(2009b)]{lbas} Abdo, A.~A., et al.\ (\Fermi-LAT 
collaboration) 2009b, \apj, 700, 597 (LAT Bright AGN Sample)

\bibitem[Abdo et al.(2010a)]{1fgl} Abdo, A.~A., et al.\ (\Fermi-LAT 
collaboration) 2010a, \apjs, 188, 405 (1FGL Catalog)

\bibitem[Abdo et al.(2010b)]{j1836} Abdo, A.~A., et al.\ (\Fermi-LAT
collaboration) 2010b, \apj, 712, 1209 (PSR J1836+5925)

\bibitem[Abdo et al.(2010c)]{psrcat} Abdo, A.~A., et al.\ (\Fermi-LAT 
collaboration) 2010c, \apjs, 187, 460 (1st Pulsar Catalog)

\bibitem[Abdo et al.(2010d)]{lbasspectra} Abdo, A.~A., et al.\ 
(\Fermi-LAT collaboration) 2010d, \apj, 710, 1271 (LBAS Spectra)

\bibitem[Abdo et al.(2010e)]{cena} Abdo, A.~A., et al.\ (\Fermi-LAT 
collaboration) 2010e, Science, 328, 725 (Cen~A Lobes)

\bibitem[Ackermann et al.(2011a)]{2lac} Ackermann, M., et al.\ 
(\Fermi-LAT collaboration) 2011a, \apj, 743, 171 (2nd LAT AGN Catalog)

\bibitem[Ackermann et al.(2011b)]{radgamma} Ackermann, M., et al.\ 
(\Fermi-LAT collaboration) 2011b, \apj, 741, 30 (Radio/$\gamma$-ray 
AGN)

\bibitem[Ackermann et al.(2012)]{latseyferts} Ackermann, M., et al.\ 
(\Fermi-LAT collaboration) 2012, \apj, 747, 104 (LAT Seyferts)

\bibitem[Atwood et al.(2009)]{atw09} Atwood, W.~B., et al.\ (\Fermi-LAT 
collaboration) 2009, \apj, 697, 1071

\bibitem[Broos et al.(2010)]{bro10} Broos, P.~S., et al.\ 2010, \apj,
714, 1582

\bibitem[Burrows et al.(2005)]{bur05} Burrows, D.~N., et al.\ 2005, 
\ssr, 120, 165

\bibitem[Canosa et al.(1999)]{can99} Canosa, C.~M., et al.\ 1999, 
\mnras, 310, 30

\bibitem[Casandjian \& Grenier(2008)]{cas08} Casandjian, J.-M., \& 
Grenier, I.~A.\ 2008, \aap, 489, 849

\bibitem[Cheung(2007)]{che07} Cheung, C.~C.\ 2007, The First GLAST 
Symposium, 921, 325

\bibitem[Cognard et al.(2011)]{cog11} Cognard, I., et al.\ 2011, \apj, 
732, 47

\bibitem[Condon et al.(1998)]{con98} Condon, J.~J., et al.\ 1998, \aj, 
115, 1693

\bibitem[Crawford et al.(2006)]{cra06} Crawford, F., et al.\ 2006, \apj, 
652, 1499

\bibitem[Cutri et al.(2012)]{cut12} Cutri, R.~M., et al.\ 2012, 
VizieR Online Data Catalog, 2307, 0

\bibitem[Falcone et al.(2011)]{fal11} Falcone, A., et al.\ 2011, 
AAS/High Energy Astrophysics Division, 12, \#04.03

\bibitem[Fruscione et al.(2006)]{fru06} Fruscione, A., et al.\ 2006, 
Proc.~SPIE, 6270, 60

\bibitem[Gehrels et al.(2004)]{geh04} Gehrels, N., et al.\ 2004, \apj, 
611, 1005

\bibitem[Halpern et al.(2002)]{hal02} Halpern, J.~P., et al.\ 2002, 
\apjl, 573, L41

\bibitem[Hartman et al.(1999)]{har99} Hartman, R.~C., et al.\ 1999, 
\apjs, 123, 79

\bibitem[Kalberla et al.(2005)]{kal05} Kalberla, P.~M.~W., et al.\ 2005, 
\aap, 440, 775

\bibitem[Keith et al.(2011)]{kei11} Keith, M., et al.\ 2011, \mnras, 
414, 1292

\bibitem[Kong et al.(2012)]{kon12} Kong, A.~K.~H., et al.\ 2012, \apjl, 
747, L3

\bibitem[La Palombara et al.(2006)]{lap06} La Palombara, N., et al.\ 
2006, \aap, 458, 245

\bibitem[Lenain et al.(2010)]{len10} Lenain, J.-P., Ricci, C., T{\"u}rler, 
M., Dorner, D., \& Walter, R.\ 2010, \aap, 524, A72

\bibitem[Maeda et al.(2011)]{mae11} Maeda, K., et al.\ 2011, \apj, 729, 
103

\bibitem[Marelli et al.(2011)]{mar11} Marelli, M., De Luca, A., \& 
Caraveo, P.~A.\ 2011, \apj, 733, 82

\bibitem[McConville et al.(2011)]{mcc11} McConville, W., et al.\ 2011, 
\apj, 738, 148

\bibitem[Mirabal et al.(2000)]{mir00} Mirabal, N., et al.\ 2000, \apj, 
541, 180

\bibitem[Monet et al.(2003)]{mon03} Monet, D.~G., et al.\ 2003, \aj, 
125, 984

\bibitem[Mukherjee \& Halpern(2004)]{muk04} Mukherjee, R., Halpern, J.\ 
2004, in Cosmic $\gamma$-ray Sources, Eds. K.~S.~Cheng \& G.~E.~Romero, 
304, 311

\bibitem[Nolan et al.(2003)]{nol03} Nolan, P.~L., Tompkins, W.~F., 
Grenier, I.~A., \& Michelson, P.~F.\ 2003, \apj, 597, 615

\bibitem[Nolan et al.(2012)]{2fgl} Nolan, P.~L., et al.\ (\Fermi-LAT
collaboration) 2012, \apjs, 199, 31 (2FGL Catalog)

\bibitem[Paredes et al.(2008)]{par08} Paredes, J.~M., et al.\ 2008, 
\aap, 482, 247

\bibitem[Plotkin et al.(2008)]{plo08} Plotkin, R., et al.\ 2008, \aj, 
135, 2453

\bibitem[Ransom et al.(2011)]{ran11} Ransom, S.~M., et al.\ 2011, \apjl, 
727, L16

\bibitem[Romani \& Shaw(2011)]{rom11} Romani, R.~W., \& Shaw, M.~S.\ 
2011, \apjl, 743, L26

\bibitem[Roming et al.(2005)]{rom05} Roming, P.~W.~A., et al.\ 2005, 
\ssr, 120, 95

\bibitem[Saz Parkinson et al.(2010)]{saz10} Saz Parkinson, P.~M., et 
al.\ 2010, \apj, 725, 571

\bibitem[Skrutskie et al.(2006)]{skr06} Skrutskie, M.~F., et al.\ 
2006, \aj, 131, 1163

\bibitem[Swanenburg et al.(1981)]{swa81} Swanenburg, B.~N., et al.\ 
1981, \apjl, 243, L69

\bibitem[Takahashi et al.(2012)]{tak12a} Takahashi, Y., et al.\ 2012, 
\apj, 747, 64

\bibitem[Takeuchi et al.(2012)]{tak12b} Takeuchi, Y., et al.\ 2012, \apj, 
749, 66

\bibitem[Tavani et al.(1997)]{tav97} Tavani, M., et al.\ 1997, \apjl, 
479, L109

\bibitem[Teng et al.(2011)]{ten11} Teng, S.~H., Mushotzky, R.~F., Sambruna, 
R.~M., Davis, D.~S., \& Reynolds, C.~S.\ 2011, \apj, 742, 66

\bibitem[Voges et al.(1999)]{vog99} Voges, W., et al.\ 1999, \aap, 349, 
389

\bibitem[Voges et al.(2000)]{vog00} Voges, W., et al.\ 2000, \iaucirc, 
7432, 3

\bibitem[Wolff et al.(2010)]{wol10} Wolff, M.~T., et al.\ 2010, BAAS, 
42, 669

\bibitem[Wright et al.(2010)]{wri10} Wright, E.~L., et al.\ 2010, \aj, 
140, 1868

\end{thebibliography}
\end{document}